\documentclass[sigconf,nonacm]{acmart}

\setcopyright{none}                      
\acmDOI{}                                 
\acmISBN{}                                
\setcopyright{none}
\renewcommand\footnotetextcopyrightpermission[1]{}
\usepackage{graphicx} 
\usepackage{enumitem}

\usepackage[utf8]{inputenc}
\usepackage{graphicx}
\usepackage[most]{tcolorbox}

\definecolor{olive}{rgb}{0.33,0.42,0.18}

\usepackage[utf8]{inputenc}
\usepackage{subcaption}
\definecolor{dimgrey}{RGB}{105,105,105}   
\usepackage{fancybox}
\usepackage{listings}
\usepackage{xcolor}
\usepackage{url}
\usepackage{xspace}
\usepackage{setspace}
\usepackage{subcaption} 
\usepackage[htt]{hyphenat}
\usepackage{todonotes}
\usepackage{booktabs}
\usepackage{threeparttable}
\usepackage{siunitx}
\definecolor{backcolour}{rgb}{0.95,0.95,0.92}
\definecolor{backcolour}{rgb}{0.95,0.95,0.92}
\definecolor{codegreen}{rgb}{0,0.6,0}
\definecolor{codegray}{rgb}{0.5,0.5,0.5}
\definecolor{codepurple}{rgb}{0.58,0,0.82}
\definecolor{codenavy}{rgb}{0.1,0.1,0.44}
\definecolor{codered}{rgb}{0.6,0,0}

\newcommand{\NA}{\multicolumn{1}{c}{--}}
\lstdefinestyle{mystyle}{
    backgroundcolor=\color{backcolour},
    frame=lines,
    framesep=1mm,
    basicstyle=\footnotesize\ttfamily,
    breaklines=true,
    columns=fullflexible,
    captionpos=b
}

\newcommand{\VideoGameBunny}{\textcolor{dimgrey}{\textsc{VideoGameBunny}}\xspace}

\title{Automated Bug Frame Retrieval from Gameplay Videos Using Vision-Language Models}
\author{Wentao Lu}
\email{wlu4@ualberta.ca}
\affiliation{%
  \institution{University of Alberta}
  \city{Edmonton}
  \state{Alberta}
  \country{Canada}
}

\author{Alexander Senchenko}
\email{asenchenko@ea.com}
\affiliation{%
  \institution{Electronic Arts}
  \city{Vancouver}
  \state{British Columbia}
  \country{Canada}
}

\author{Abram Hindle}
\email{abram.hindle@ualberta.ca}
\affiliation{%
  \institution{University of Alberta}
  \city{Edmonton}
  \state{Alberta}
  \country{Canada}
}

\author{Cor-Paul Bezemer}
\email{bezemer@ualberta.ca}
\affiliation{%
  \institution{University of Alberta}
  \city{Edmonton}
  \state{Alberta}
  \country{Canada}
}

\begin{document}

\newcommand{\RQframeselect}{How accurately can keyframes represent bug moments in gameplay videos?}
\newcommand{\RQhowwell}{How accurately can a VLM retrieve bug frames from a set of keyframes given a bug description?}

\newcommand{\RQtype}{How does a VLM handle different categories of visual bugs and what are the limitations?}

\newcommand{\RQconfidence}{How consistent is bug frame retrieval by a VLM across repeated runs?}

\begin{abstract}
\label{sec:abstract}

 Modern game studios deliver new builds and patches at a rapid pace, generating thousands of bug reports, many of which embed gameplay videos. To verify and triage these bug reports, developers must watch the submitted videos. This manual review is labour-intensive, slow, and hard to scale. In this paper, we introduce an automated pipeline that reduces each video to a single frame that best matches the reported bug description, giving developers instant visual evidence that pinpoints the bug.

Our pipeline begins with \textit{FFmpeg} for keyframe extraction, reducing each video to a median of just 1.90\% of its original frames while still capturing bug moments in 98.79\% of cases. These keyframes are then evaluated by a vision–language model (GPT-4o), which ranks them based on how well they match the textual bug description and selects the most representative frame. We evaluated this approach using real-world developer-submitted gameplay videos and JIRA bug reports from a popular First-Person Shooter (FPS) game. The pipeline achieves an overall F1 score of 0.79 and Accuracy of 0.89 for the top-1 retrieved frame. Performance is highest for the Lighting \& Shadow (F1 = 0.94), Physics \& Collision (0.86), and UI \& HUD (0.83) bug categories, and lowest for Animation \& VFX (0.51).

By replacing video viewing with an immediately informative image, our approach dramatically reduces manual effort and speeds up triage and regression checks, offering practical benefits to quality assurance (QA) teams and developers across the game industry.
\end{abstract}



\maketitle

\begin{figure*}[ht!]
  \centering
  \begin{tcolorbox}[
      colback=white,
      colframe=black,
      width=\textwidth,
      title=\bfseries Bug frames retrieved in gameplay video by GPT-4o,
      fonttitle=\bfseries,
      center title,
      boxrule=0.8pt,
      arc=2pt,
      outer arc=2pt,
      boxsep=4pt,
      left=4pt,
      right=4pt,
      top=4pt,
      bottom=4pt
    ]
    \centering
    \begin{tabular}{ccccc}
    $\ldots$ &
      \includegraphics[width=0.28\textwidth]{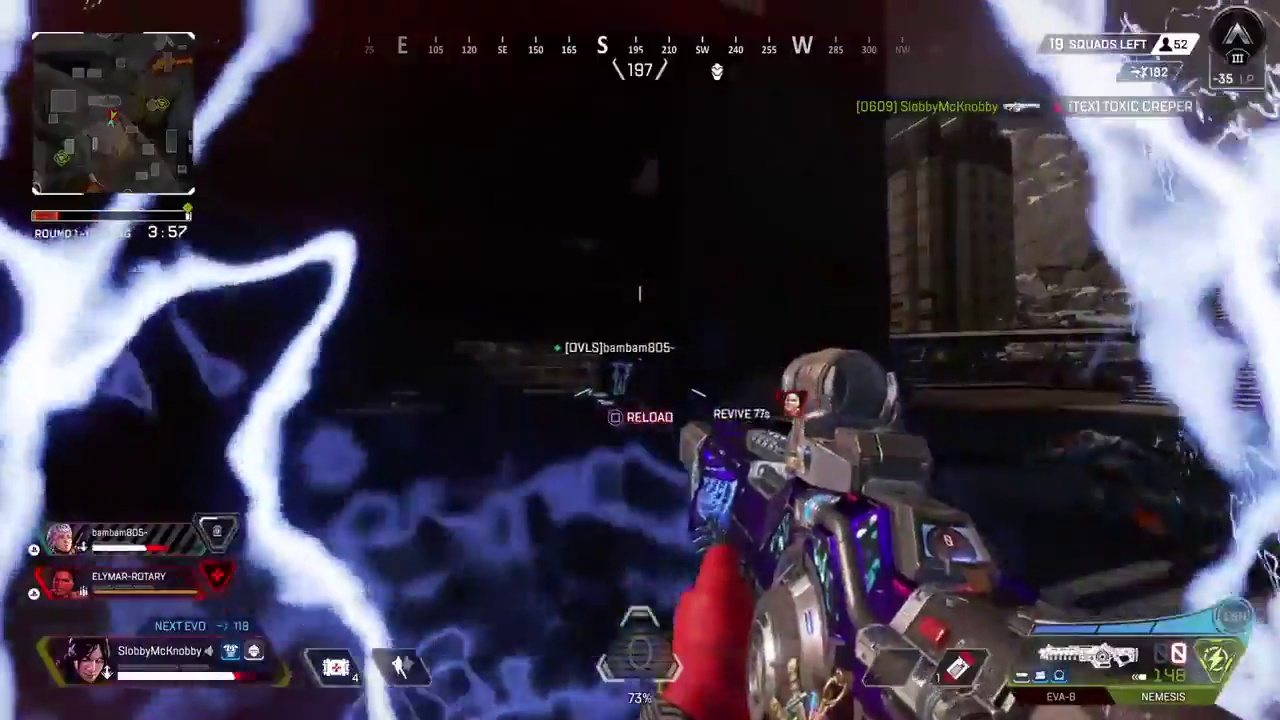} &
      \includegraphics[width=0.28\textwidth]{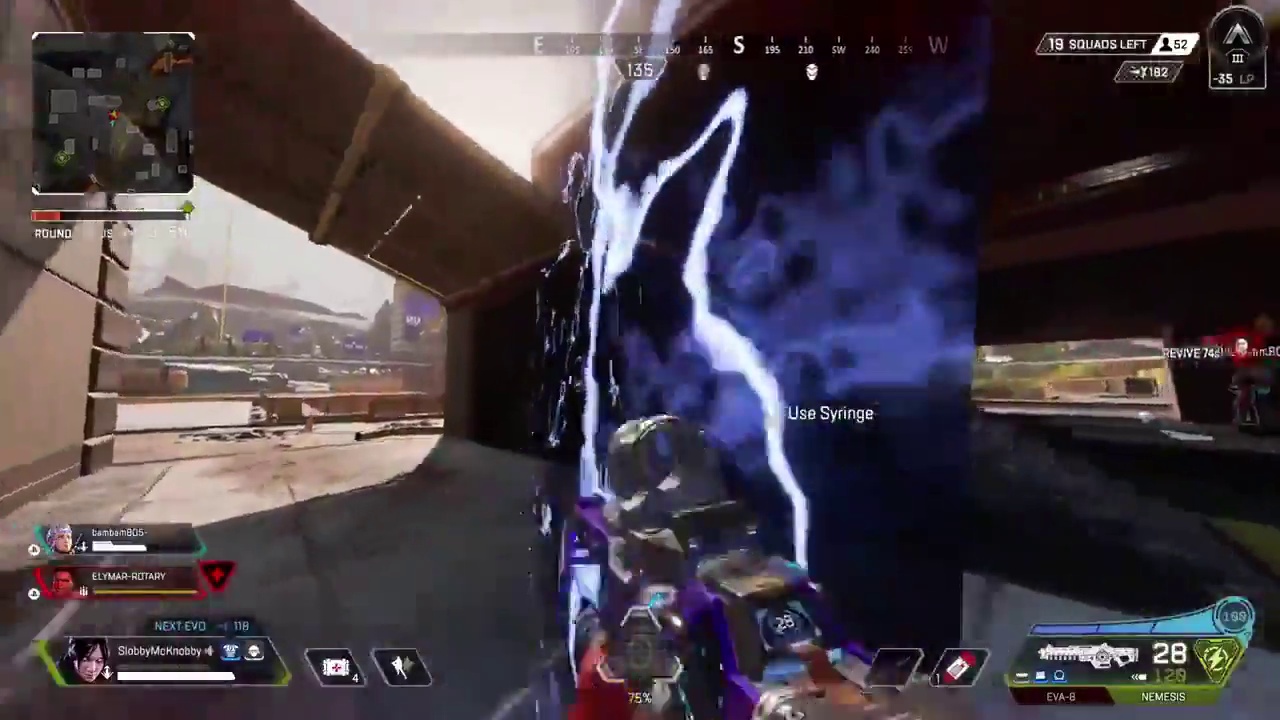} &
      $\ldots$ 
      \includegraphics[width=0.28\textwidth]{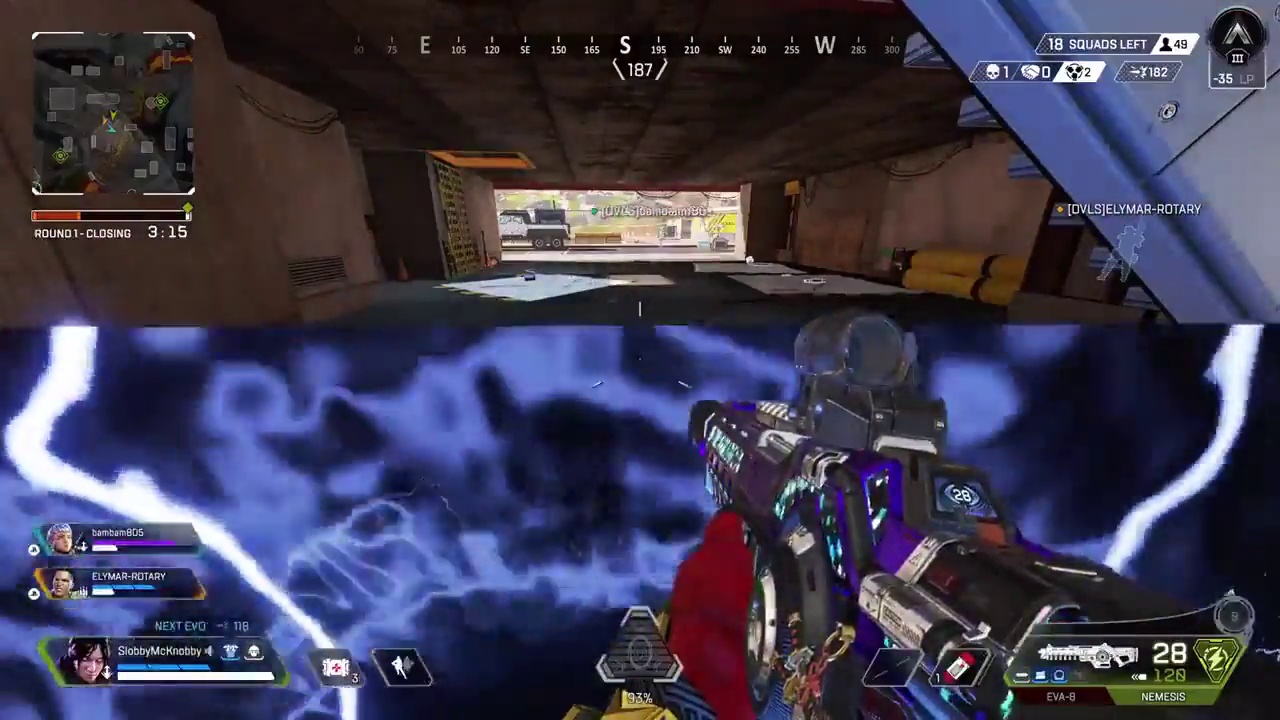} \\
      &{\scriptsize\texttt{000030.jpg}} &
      {\scriptsize\texttt{000031.jpg}} &
      {\scriptsize\texttt{000075.jpg}} &\\
    \end{tabular}

    \vspace{1ex}

    \noindent
    \textbf{\textcolor{blue}{Prompt:}} Find the frame that matches the bug description.\\[0.5ex]
    \textbf{\textcolor{olive}{Bug description:}} Visual bug when Wraith finisher gets interrupted by the portal animation.\\[0.5ex]
    \textbf{\textcolor{red}{GPT-4o:}} Based on the provided frames and prior explanations, frames 000031.jpg, 000032.jpg, and 000065.jpg clearly show the gameplay scenario where the portal animation appears clipped and blocks the player's view, and the visual evidence of the bug is most likely observable. These frames are ranked in descending order of clarity and relevance to the bug description.
  \end{tcolorbox}

  \caption{Examples of bug frames identified and retrieved in gameplay video sourced from reddit~\cite{reddit_video_example}. The prompt is for demonstration only; the actual prompt design is discussed in a later section.}
  \label{fig:gpt4o-example}
\end{figure*}

\section{Introduction}
\label{Sec:intro}

The rapid evolution of digital gaming has led to video games becoming increasingly complex software systems. As developers and testers engage with these systems, they encounter and report thousands of bugs. Among these, visual bugs can be particularly impactful, as they can significantly degrade the player experience. Prior work in traditional software development has shown that reviewing large volumes of bug reports is tedious and time-consuming~\cite {azizi2023automaticbugdetectiongames, yan2024semanticguiscenelearning,guglielmi2022usinggameplayvideosdetecting, Senchenko_2022, cooper2021takestangocombiningvisual}.
For games, this challenge is further amplified by the inherently visual nature of bugs. Reports often include gameplay videos to convey issues that are difficult to describe in text alone. For example, visual bugs such as missing textures, lighting glitches, or character clipping are better presented in a visual manner. 

Prior studies have shown that augmenting bug reports with screenshots~\cite{tan2025imagerenhancingbugreport,whatmakesgoodbugreport, wang2025empiricalstudyleveragingimages}, animated GIFs~\cite{feng2022gifdroidautomatedreplayvisual}, or videos~\cite{cooper2021takestangocombiningvisual, Moran_2015} can improve reproducibility and reduce fix times. The added visual context enables developers to more quickly understand, confirm, and prioritize issues. However, this benefit comes at the cost of increased review time, as developers are still required to watch the visual contents.

Recent advances in computer vision and natural language processing have enabled new methods for automated video understanding. In particular, vision–language models (VLMs), such as GPT-4o, Claude, Gemini, and LLaVA, can jointly interpret visual inputs and textual prompts~\cite{li2024llava, GPT4O, Claude, Gemini2.0}. These models show strong potential to automate tasks traditionally performed by humans, including identifying visual anomalies and verifying gameplay bugs~\cite{taesiri2025videogameqabenchevaluatingvisionlanguagemodels, Taesiri_VideoGameBunny, taesiri2022clipmeetsgamephysicsbug}.

In this study, we present a novel pipeline (shown in Figure~\ref{fig:gpt4o-example}) that leverages a vision-language model to assist with the triage of bug reports that contain gameplay videos. Our approach reduces each bug report video to a single, automatically selected frame that best represents the described bug, enabling triagers to quickly confirm a reported bug. Our approach leverages:
\begin{itemize}
    \item \textbf{Keyframes} (snapshots in gameplay videos that capture major visual changes) are used as a down-sampling method to extract important frames from videos. These keyframes are inserted by video encoders and often coincide with major scene transitions, either based on abrupt visual changes (e.g., rendering glitches, camera shifts) or fixed intervals~\cite{FFmpeg, korolkov2025scenedetectionpolicieskeyframe, effective_keyframe_extraction}. This makes them a natural choice for gameplay analysis, allowing us to reduce the video to only 1.90\% of its frames while preserving most visually distinct moments.
    \item A \textbf{Vision-Language Model} (GPT-4o~\cite{GPT4O}) ranks the extracted keyframes against the bug description and returns the single frame most likely to depict the reported problem.
\end{itemize}

We focus on the following research questions:

\textbf{RQ1: \RQframeselect} 
To ensure that keyframes can indeed capture bug moments, we evaluated keyframe-based extraction using \textit{FFmpeg}~\cite{FFmpeg}. 
We found that this approach retained at least one bug frame in 98.79\% of videos while sampling only 1.90\% of all frames (with approximately 10 keyframes per video and an average of 4.79 keyframes that represent the bug). Thus, keyframe extraction offers an efficient down-sampling strategy that still preserves the most representative frames of a visual bug.

\textbf{RQ2: \RQhowwell} 
A retrieval system must reliably extract the relevant frame that captures the bug from this potentially noisy and varied input. We evaluated our pipeline on 350 videos and achieved an overall 89\% Accuracy@1 and F1@1 score of 0.79 for the top returned frame. The pipeline performs best on Lighting \& Shadow (F1 = 0.94), Physics \& Collision (0.86), and UI \& HUD (0.83). This shows that the VLM can usually retrieve a representative bug frame and thus greatly reduce the effort developers spend reviewing video footage.


\textbf{RQ3: \RQconfidence}  
Because of the inherent non-determinism of VLMs, it is important to evaluate the consistency of bug frame retrieval across repeated runs.  Although the overall F1 score remained stable across runs (median = 0.79), variation in true and false positives highlights the non-deterministic nature of VLM outputs. For 183 out of 350 bug videos, the retrieval outcome was consistently correct across all runs. This number increases to 215 when applying a simple majority vote over 3 runs.

The main contributions of our paper are as follows.
\begin{itemize}
    \item We developed an automated pipeline that integrates VLM capabilities (GPT-4o) with video frame analysis to return a single frame that matches a textual bug description.
    \item We show that keyframes from gameplay videos preserve bug-relevant content and can be used as a source of bug frame retrieval.
    \item We quantify the pipeline’s overall accuracy, category-specific performance, and run-to-run stability on real bug reports from an 
    AAA game, providing a comprehensive assessment of its performance and reliability in bug frame retrieval.
\end{itemize}

\textbf{Paper organization:} Section~\ref{sec:related_work} reviews related work. Section~\ref{sec:methodology} introduces our pipeline, including keyframe extraction and VLM usage, broken down into four steps. Section~\ref{sec:exp_setup} details the experimental setup. Section~\ref{sec:results} answers the research questions. Section~\ref{sec:discussion} compares our results with prior work and discusses the limitations of our approach. Section~\ref{Sec:threats} discusses threats to validity, and Section~\ref{sec:conclu} concludes our work.

\section{Related Work}
\label{sec:related_work}

Driven by advancements in both data collection (automated gameplay and streamers' footage)~\cite{auto_game_test, CV_auto_testing, Replication_Game_Debugging} and analysis techniques, including recent progress in bug detection and retrieval~\cite{taesiri2022clipmeetsgamephysicsbug,taesiri2025videogameqabenchevaluatingvisionlanguagemodels, Taesiri_VideoGameBunny, truelove2023findingneedlehaystackdetecting}, gameplay videos are increasingly important in game quality assurance.
In this section, we review work in three primary areas: (1) traditional quality assurance in games, (2) bug detection from gameplay videos, and (3) VLM/LLM assistants for video game QA.

\subsection{Traditional quality assurance in games}
Game quality assurance (QA) traditionally relies heavily on manual efforts, including playtesting, visual inspection, and reproduction of reported issues by dedicated QA personnel~\cite{guglielmi2022usinggameplayvideosdetecting, Senchenko_2022, cooper2021takestangocombiningvisual}. While effective for identifying critical failures such as game crashes, these workflows are highly labor-intensive and time-consuming~\cite{azizi2023automaticbugdetectiongames, zhao2021lightweightapproachhumanlikeplaytesting, Senchenko_2022,cooper2021takestangocombiningvisual}. In particular, bug reports often include video recordings as supporting evidence, each of which requires manual review and interpretation. Furthermore, the vast number of possible game states and the non-determinism introduced by factors such as multithreading and randomness make exhaustive manual QA impractical for modern games~\cite{guglielmi2022usinggameplayvideosdetecting}. These challenges have motivated growing interest in automation and tooling to assist various aspects of the QA process.

\textbf{Our work addresses a specific pain point within this landscape} by automating the retrieval of visually relevant frames from gameplay videos associated with bug reports. Rather than replacing testers, our method aims to support QA workflows by surfacing candidate frames for inspection, reducing the need for exhaustive manual scrubbing of video footage.

\subsection{Bug detection in gameplay videos}
Early studies explored the potential of mining gameplay videos to extract information about bugs and glitches. One of the first works in this space is by Lin et al.~\cite{Dayi_game_bug_video}, who investigated whether gameplay videos that showcase bug videos can be automatically identified using publicly available metadata from online platforms with a precision of 0.80. Building on this idea of mining gameplay content, Guglielmi et al.~\cite{guglielmi2022usinggameplayvideosdetecting,guglielmi2023usinggameplayvideosdetecting} introduced an approach to automatically extract relevant segments from gameplay videos by partitioning videos into meaningful units based on streamer captions, shifting the timestamp by $k$ seconds before the video cuts, and categorizing the detected anomalies by type and context through Machine Learning (ML). Similarly, TrueLove et al.~\cite{truelove2023findingneedlehaystackdetecting} proposed an ML-based method to classify video segments, also split based on video transcript, as buggy or non-buggy, and analyzed visual features that correlate with bug occurrences. Guglielmi et al.~\cite{Stuttering} introduced \textit{HASTE}, a tool designed to detect stuttering events in gameplay videos. Azizi et al.~\cite{azizi2023automaticbugdetectiongames} framed bug detection as an anomaly detection problem and leveraged Long-Short-Term Memory (LSTM) networks to identify perceptual and behavioral bugs in gameplay videos. By modeling temporal dependencies in video sequences, their framework is able to flag unusual frames without relying solely on rule-based methods. 

These methods rely heavily on captions or transcripts for video segmentation. This introduces key limitations: (1) such captions do not exist in developer-submitted gameplay footage, and (2) the timing of transcripts may not align precisely with the occurrence of the visual bug, resulting in incomplete or misaligned segment boundaries. Moreover, some techniques focus only on certain bug types (e.g. stuttering), limiting their applicability to the broader range of visual bugs encountered in game development.



While prior methods aim to detect visual bugs through anomaly classification or transcript-driven segmentation, \textbf{our goal is fundamentally different: we focus on retrieving visually relevant frames from gameplay videos based on natural language bug descriptions}. To support this, we introduce a deterministic keyframe-based segmentation strategy that enables frame-level analysis without requiring captions or transcripts. 
Unlike prior work that often relies on specialized content from streamers or online platforms, our evaluation is conducted on industrial gameplay data, making our findings directly applicable to real-world QA workflows.






\subsection{VLM/LLM assistants for video game QA}

Beyond bug detection, there is growing interest in employing large multimodal models as vision assistants in the video game domain~\cite{hu2025surveylargelanguagemodelbased, ma2024largelanguagemodelsplay, Roadmap_LLM_game}. Taesiri et al.~\cite{taesiri2022clipmeetsgamephysicsbug} show promising results by leveraging zero-shot transfer capabilities of the Contrastive Language Image Pre-Training (CLIP) model to query bug images. They also introduced \VideoGameBunny~\cite{Taesiri_VideoGameBunny}, a model fine-tuned on extensive video game image datasets and associated instruction pairs, demonstrating an improved understanding of the game context and more accurate in-game responses. 
In subsequent work, Taesiri et al.~\cite{taesiri2025videogameqabenchevaluatingvisionlanguagemodels, taesiri2024glitchbenchlargemultimodalmodels} proposed several benchmark suites, including \textit{VideoGameQA-Bench} and \textit{GlitchBench}, to evaluate the capabilities of state-of-the-art vision-language models across a range of video game QA tasks. These benchmarks assess model performance on tasks such as glitch detection in both static images and gameplay videos, highlighting the potential and current limitations of using large multimodal models for bug understanding/detection in games.

Melhart et al.~\cite{melhart2025largelanguagemodelscapture} investigate whether large language models can capture continuous player engagement from gameplay videos. Their comprehensive evaluation examines state-of-the-art models, including variants of GPT-4o~\cite{GPT4O} and LLaVA~\cite{li2024llava}, using one-shot and few-shot multimodal prompting strategies. Although the study reports that the overall accuracy of LLM-based engagement prediction only marginally surpasses a majority class baseline, it highlights that model performance is highly sensitive to input modality, prompting strategy, and even the specific context of the game. These insights not only underscore the promise of integrating LLMs into game content analysis but also reveal current limitations that motivate further research in autonomous gameplay video interpretation.


While prior work has applied vision-language models to QA tasks such as glitch classification or engagement prediction, \textbf{our method explores a different application: using VLMs for frame-level retrieval guided by natural language bug descriptions.} Specifically, we use GPT-4o in a zero-shot setting~\cite{zero_shot_Vlm, meta_prompting_zero_shot} to rank keyframes based on their semantic alignment with the bug report description. Additionally, we evaluate the stability of this retrieval process via multi-run consistency checks, a dimension not explored in previous studies.

\begin{figure*}[t!]
    \centering
    \includegraphics[width=0.96\textwidth]{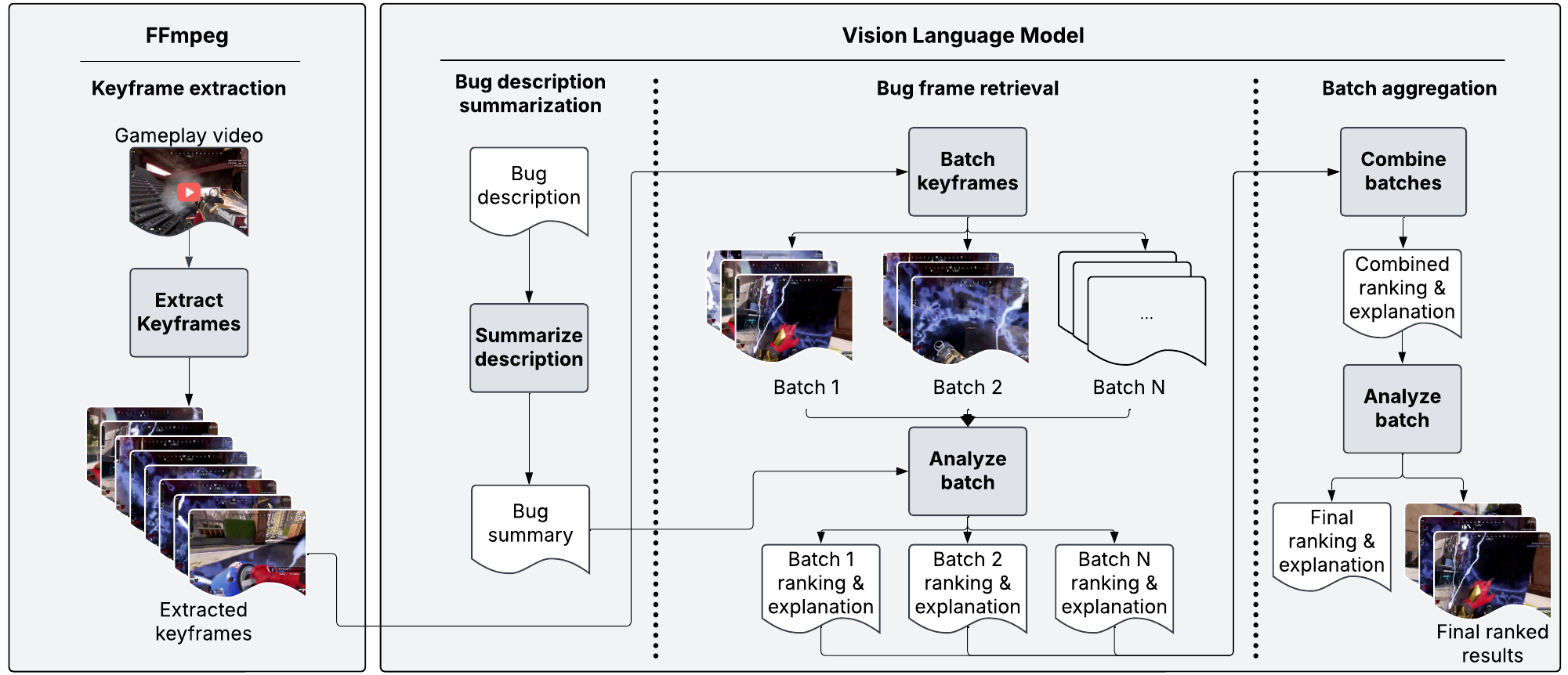}
    \caption{Overview diagram of our methodology (example video sourced from Reddit~\cite{reddit_video_example} as an illustration of the workflow).}
    \label{fig:flow}
\end{figure*}

\section{Methodology}
\label{sec:methodology}

This section details our approach for bug frame retrieval in gameplay videos using VLMs. The integrated pipeline consists of four parts: keyframe extraction, bug description summarization, bug frame retrieval, and batch aggregation. We provide an overview of our methodology in Figure~\ref{fig:flow}.

\subsection{Keyframe extraction}
\label{sec:keyfream_extract}

To reduce computational overhead while preserving bug moments in gameplay videos, we adopt a keyframe-based down-sampling strategy. We leverage the internal encoding structure of video files to extract visually significant moments using keyframes.

We utilize \textit{FFmpeg}~\cite{FFmpeg}, a widely adopted tool for multimedia processing, to extract keyframes. In general, the original videos uploaded by developers may be encoded using unknown settings, and keyframes are determined during video encoding based on various factors~\cite{effective_keyframe_extraction, korolkov2025scenedetectionpolicieskeyframe,H264_encoder}, including:

\begin{itemize}
\item \textbf{Scene change detection:} A sharp shift in visual content (e.g., from one scene to another) often triggers a new keyframe.
\item \textbf{Fixed interval configuration:} Encoders may insert keyframes at fixed time intervals or frame counts (e.g., every 120 frames) in cases where no significant visual change occurs.
\end{itemize}

As a result, keyframe frequency and distribution vary significantly across videos due to differences in their original encoding settings. Some videos may contain very few keyframes if encoded with long fixed intervals or without scene change detection, which can lead to uneven sampling and the potential omission of bug moments. Therefore, we re-encode videos using \textit{FFmpeg} (with the H.264 encoder under the medium preset) to enhance bug moment coverage through more informative keyframes. 
This normalization ensures that keyframes are inserted consistently across videos.

The output of this step is a set of keyframes for each gameplay video. The number of frames extracted varies, from a few to several hundred (see Section~\ref{sec:rq_keyframes}), depending on the duration and how frequently the visual content of the video changes. These frames represent a compressed visual summary of the video, from which at least one is expected to represent the bug described in the corresponding report. These keyframes are then passed to the GPT-4o model for visual interpretation in the following stages.

\subsection{Bug description summarization}
Before starting visual analysis, we first generate a concise bug description. While JIRA reports include a dedicated \texttt{description} field, in our data, developers often paste large blocks of text into it, including logs, internal tags, build information, and in-game coordinates. These details, while useful for debugging and reproducing a bug, are irrelevant to the task of visual bug frame retrieval and consume valuable input tokens when passed to the VLM.

To remove excess info in the description and reduce token usage, we use GPT-4o to generate summaries that are concise and short for downstream visual analysis tasks. Each generated summary consists of two parts: (1)~a concise restatement of the core issue described in the report, and (2)~a binary indication of whether the bug is expected to produce a visible effect in the gameplay video. The second component is critical as it allows the following stages in the pipeline to bypass further visual processing for bugs that are likely of a non-visual nature. 

To evaluate the reliability of this binary indication, one of the authors verified whether the indication correctly reflected the nature of the bug in a random sample of 100 GPT-4o-generated summaries. The model's answers showed 97\% agreement with human judgment. 

\subsection{Bug frame retrieval}
For each gameplay video, the extracted keyframes are passed to GPT-4o alongside the corresponding bug description summary. The model is instructed to perform the following tasks:
\begin{itemize}
 \item Return \texttt{none} if the bug described in the text is not visually observable in any of the frames.
 \item Identify frames that visually depict the bug based on the provided summary.
 \item Rank the identified frames in descending order of how clearly they represent the described bug.
 \end{itemize}
The model is also asked to provide an explanation for its ranking, including visual reasoning grounded in the bug description. The response is returned in a structured JSON format with two keys:
 \begin{itemize}
 \item \textit{Explanation:} A natural language summary justifying the model’s frame selection and ranking;
 \item \textit{Ranking:} A ranked list of frame filenames.
 \end{itemize}
Due to GPT-4o’s input token limit, a maximum of 50 images can be processed in a single chat completion~\cite{GPT_API}. While most gameplay videos in our dataset yield fewer than 50 keyframes, a few outliers (in our studied data, only 2 videos) exceed this threshold. To accommodate such cases, we partition the extracted keyframes into consecutive batches, preserving their original order of sequences. This sequential grouping retains contextual continuity; for instance, if a bug occurs immediately after a player action, the relevant batch is likely to include both the trigger and its visual consequence.
This batching process is applied iteratively across all extracted frames in the video. Each batch is processed independently using the same prompt, and the final ranked frame list for the video is constructed by merging the outputs across all batches. For these outlier videos, we observed that batch size had a negligible impact on retrieval performance.

\subsection{Batch aggregation}
After processing all batches, we obtain multiple sets of ranked keyframes, each accompanied by GPT-4o’s batch-level explanation. To combine these partial results and produce a final, globally consistent ranking of bug-relevant frames, we perform a re-evaluation stage, following recent advances in which large language models can be used as judges~\cite{zheng2023judgingllmasajudgemtbenchchatbot, gu2025surveyllmasajudge, li2024llmsasjudgescomprehensivesurveyllmbased, dong2024llmpersonalizedjudge}.

We aggregate all candidate frames and their batch-level explanations into a single prompt and pass it to GPT-4o for a second pass. The model is instructed to:
\begin{itemize}
 \item Re-rank all candidate frames based on how clearly they depict the bug described in the original JIRA description.
 \item Reference the batch-level explanations.
 \end{itemize}
The model’s response is returned in the same structured JSON format as above.
In the case where there is only a single batch, this step still applies to re-check only on the one batch.

\begin{figure*}[htbp]
  \centering
  \begin{subfigure}[b]{0.69\columnwidth}
    \centering
    \includegraphics[width=\linewidth]{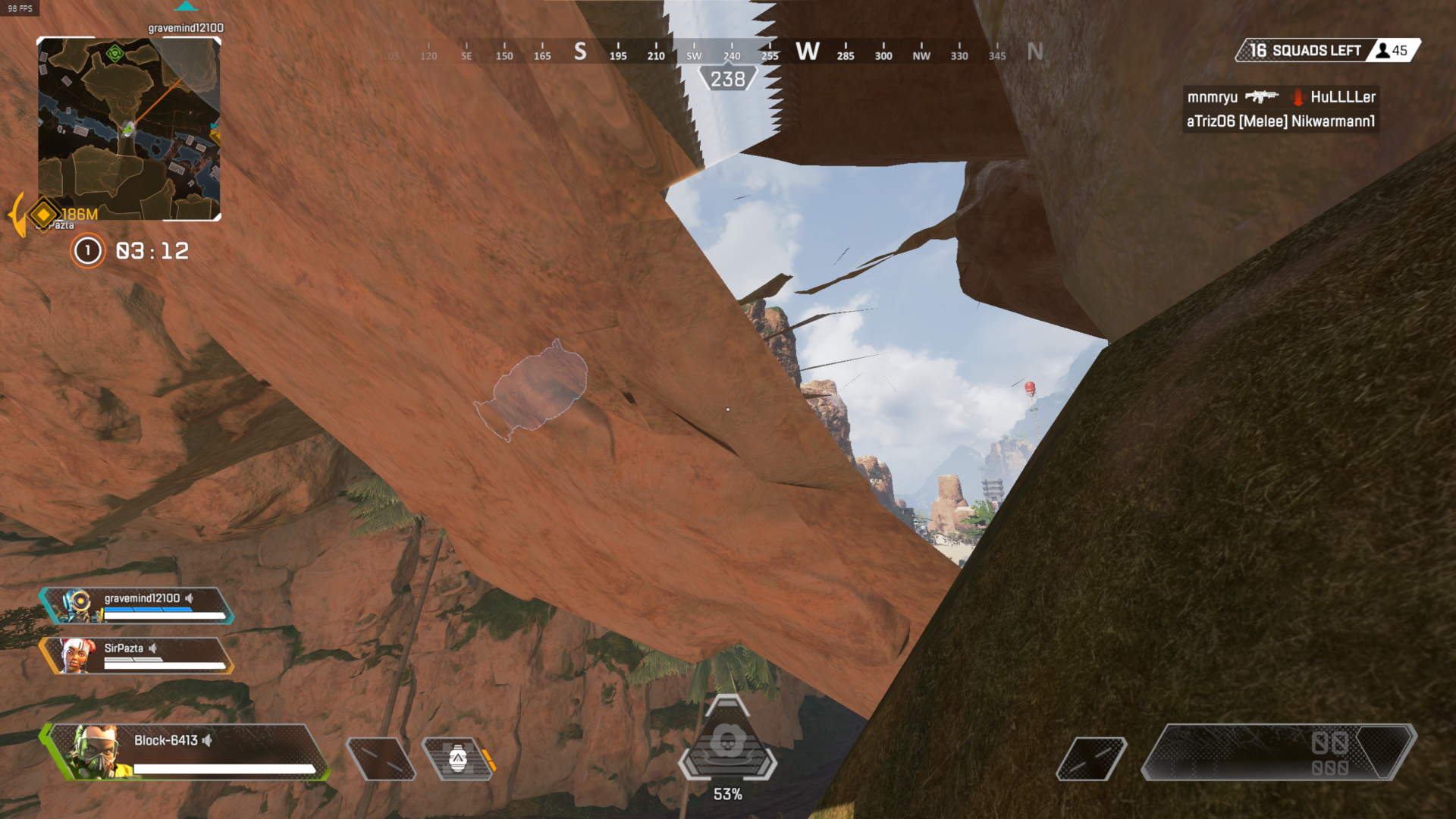}
    \caption{Player hides underneath a rock}
    \label{fig:clipping}
  \end{subfigure}\hfill
  \begin{subfigure}[b]{0.69\columnwidth}
    \centering
    \includegraphics[width=\linewidth]{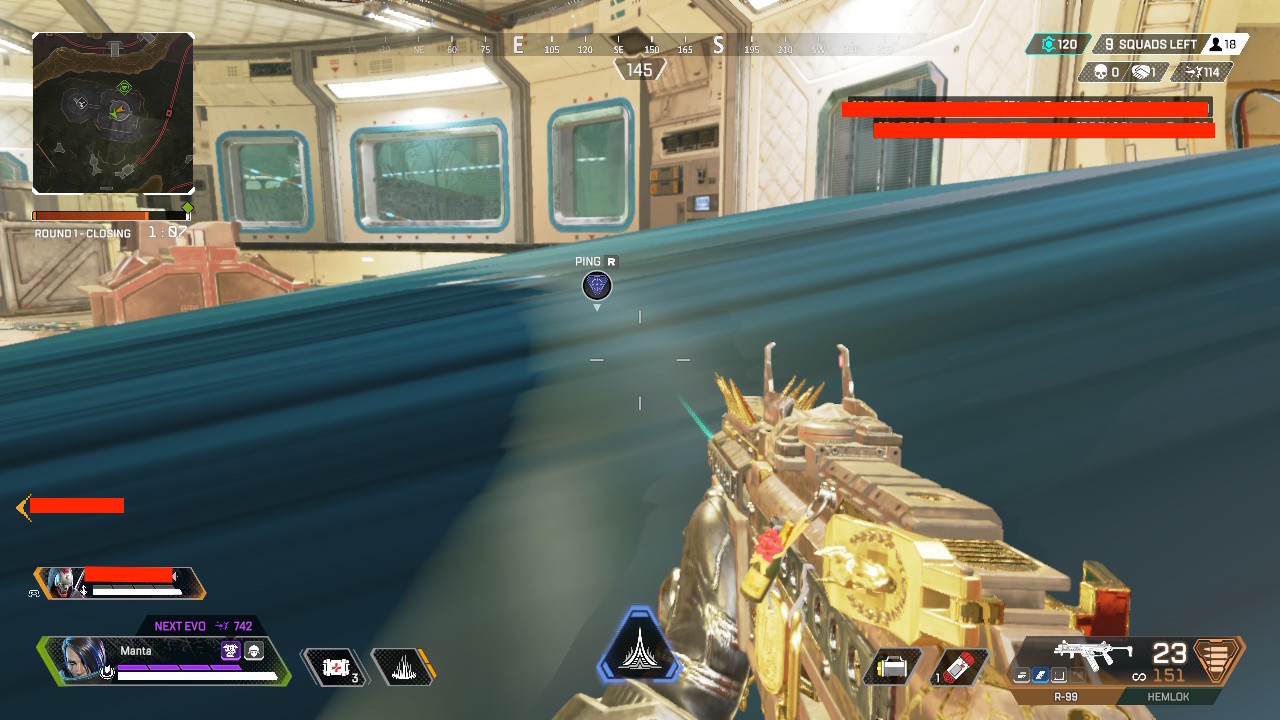}
    \caption{Stretched ground texture rendering}
    \label{fig:texture}
  \end{subfigure}\hfill
  \begin{subfigure}[b]{0.69\columnwidth}
    \centering
    \includegraphics[width=\linewidth]{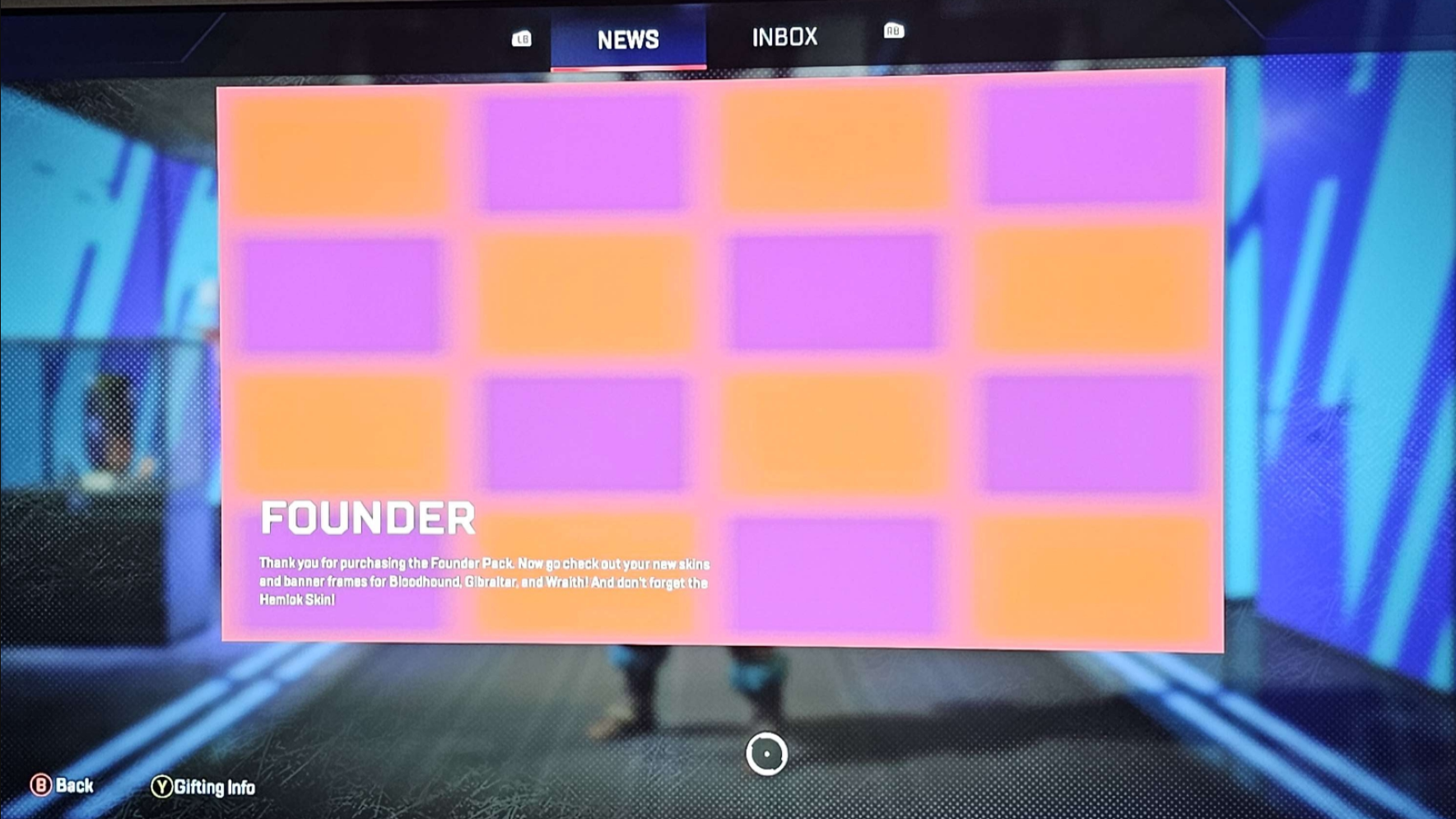}
    \caption{Placeholder UI image}
    \label{fig:ui}
  \end{subfigure}

  \vspace{1ex}

  \begin{subfigure}[b]{0.69\columnwidth}
    \centering
    \includegraphics[width=\linewidth]{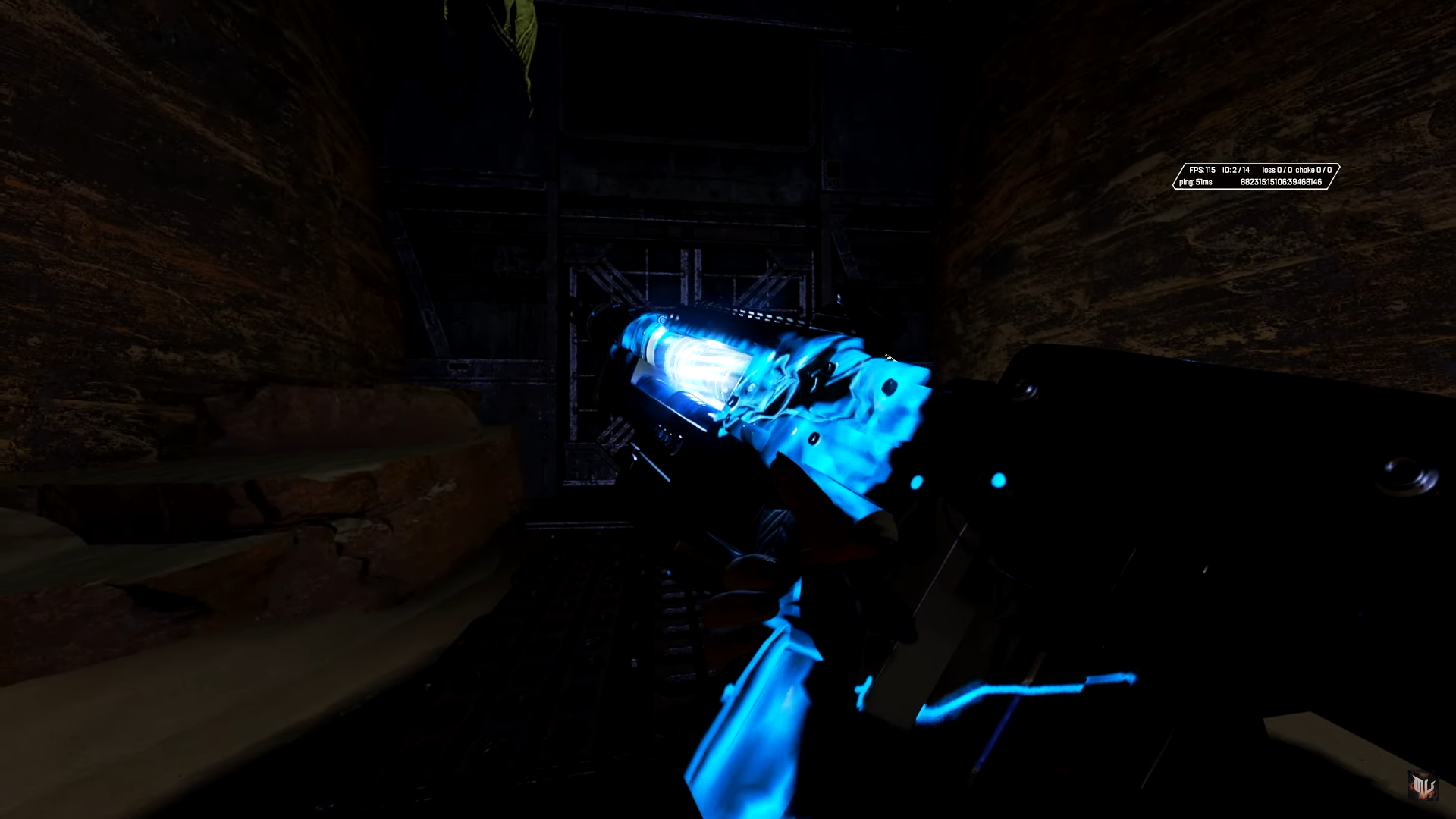}
    \caption{Scene too dark}
    \label{fig:lighting}
  \end{subfigure}\hfill
  \begin{subfigure}[b]{0.69\columnwidth}
    \centering
    \includegraphics[width=\linewidth]{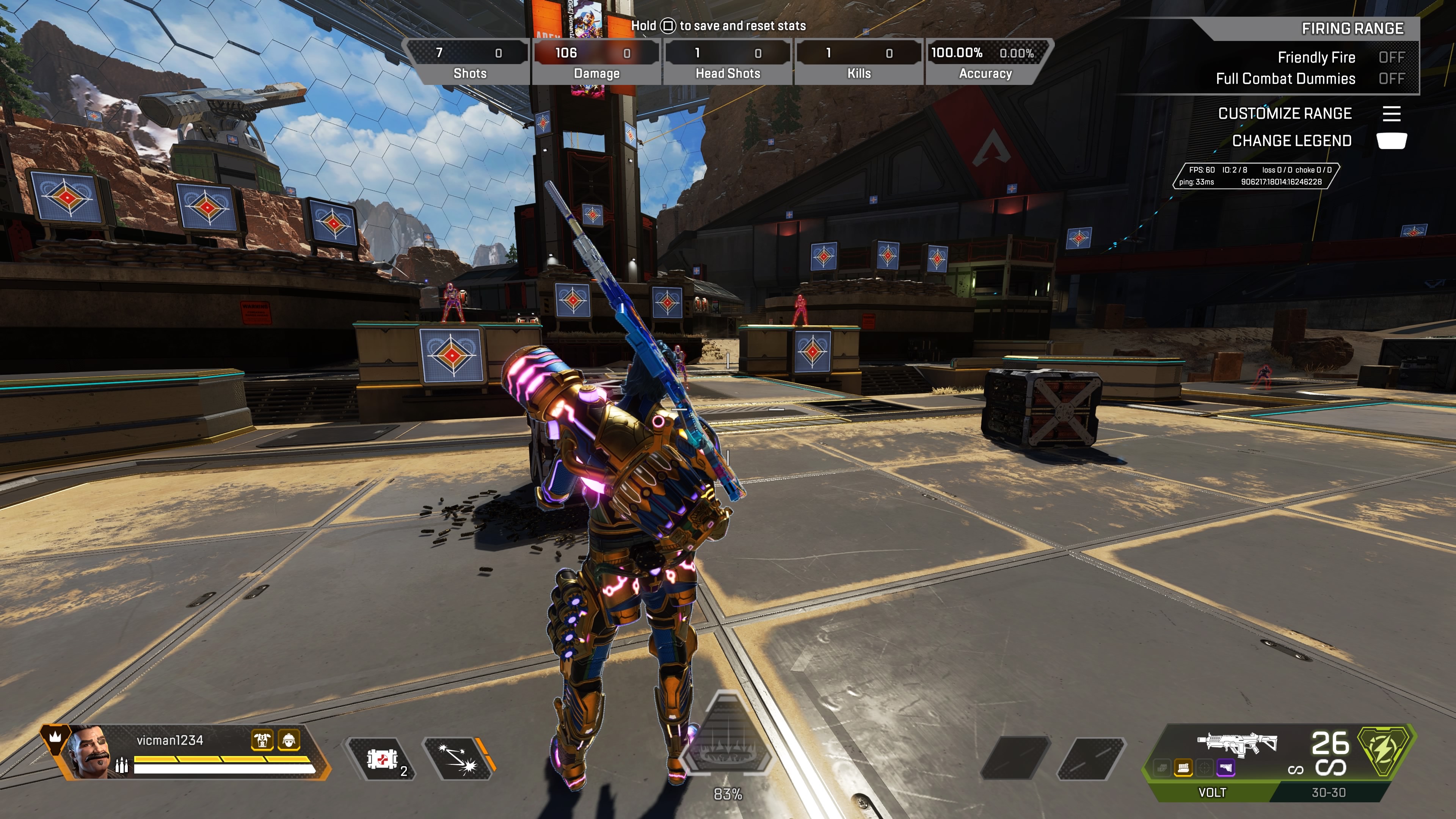}
    \caption{Weapon floats over the character's back}
    \label{fig:animation}
  \end{subfigure}\hfill
  \begin{subfigure}[b]{0.69\columnwidth}
    \centering
    \includegraphics[width=\linewidth]{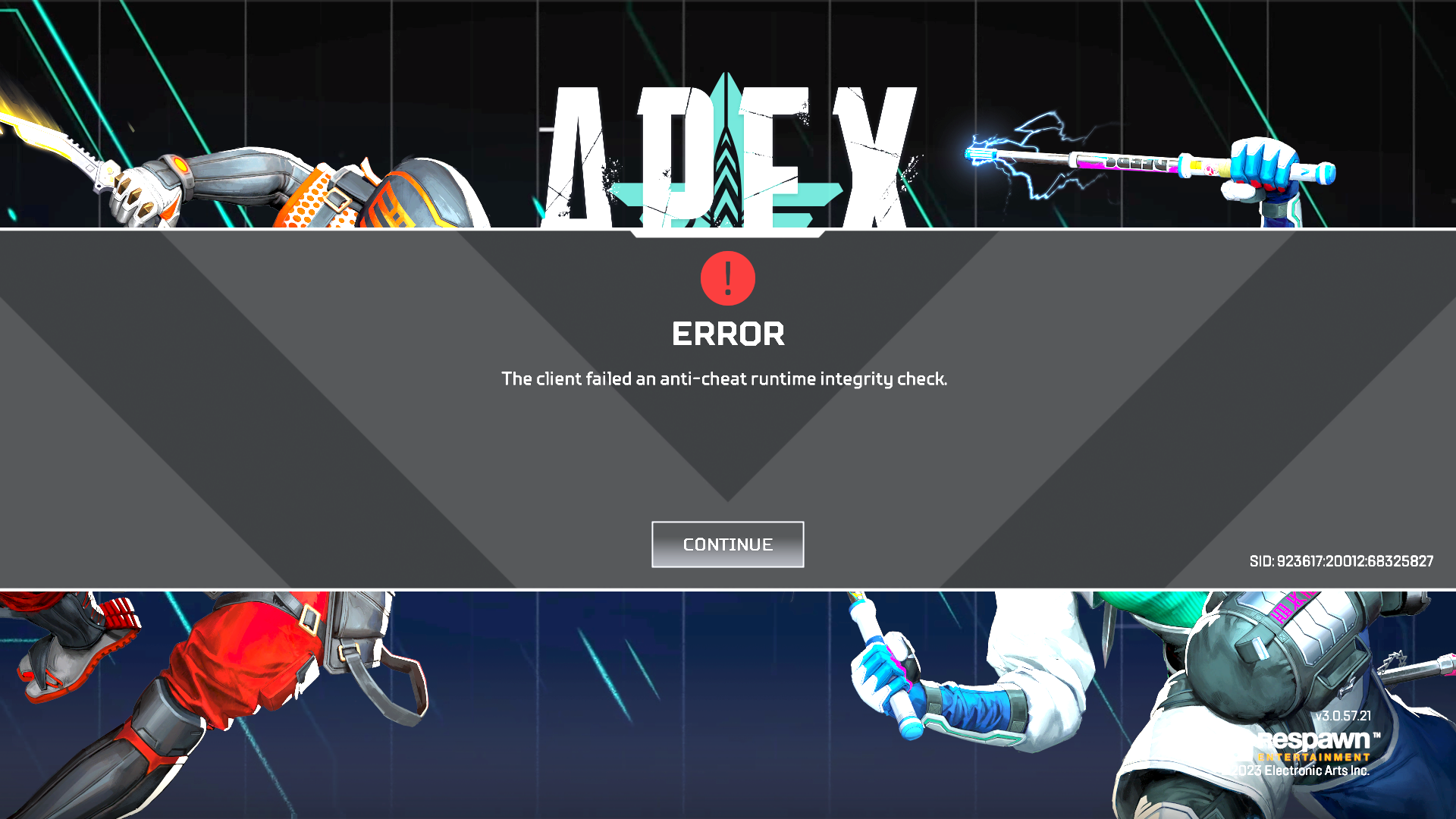}
    \caption{Game crash error}
    \label{fig:crash}
  \end{subfigure}

  \caption{Six samples of visual bugs in game videos: (a) Physics \& Collision, (b) Rendering \& Texture, (c) Camera \& UI, (d) Lighting \& Shadow, (e) Animation \& VFX, (f) Game Crash \& Logic. All images are captured from online sources~\cite{EA_forum}.}
  \label{fig:bug-types}
\end{figure*}

\begin{table*}[htbp]
  \centering
  \caption{Bug Category Definitions and Sample Counts}
  \label{tab:category-definitions}
  \begin{tabular}{@{}p{3cm}p{8cm}r@{}}
    \toprule
    \textbf{Category} & \textbf{Description} & \textbf{\#Samples} \\
    \midrule

    Physics \& Collision
      & Issues arising from the physics engine or collision detection (e.g.\ objects/players passing through geometry). 
      & 671 \\

    Animation \& VFX
      & Problems with character or object animations and visual effects (e.g.\ body distortion, particle glitches). 
      & 861 \\

    Rendering \& Texture
      & Errors in the rendering pipeline or asset texturing (e.g.\ missing textures, incorrect material artifacts). 
      & 418 \\

    UI \& HUD
      & Bugs in Heads-Up Display and User-Interface elements (e.g.\ HUD misalignment and UI images). 
      & 374 \\

    Performance
      & Frame-rate drops, stuttering, or other performance degradations under normal gameplay. 
      & 137 \\

    Lighting \& Shadow
      & Faults in scene lighting or post-process effects (e.g.\ incorrect shadows, abnormally dark environment). 
      & 66 \\

    Game Crash \& Logic
      & Game crashes, incorrect game logic flows or functionality behavior (e.g.\ error messages, player rewards tracking). 
      & 295 \\

    Unknown Category
      & \NA 
      & 58 \\
    \bottomrule
  \end{tabular}
\end{table*}

\begin{table*}[htbp]
  \centering
  \footnotesize
  \caption{Bug frame retrieval performance by category}
  \label{tab:frame-detection}
  \resizebox{0.9\textwidth}{!}{%
  \begin{threeparttable}
    \begin{tabular}{
      @{} l
      c        
      *{3}{c}  
      *{3}{c}  
      r
      *{3}{c}  
      c        
      @{}
    }
      \toprule
      & \textbf{\#Bug}
      & \multicolumn{3}{c}{\textbf{TP (Top N)}} 
      & \multicolumn{3}{c}{\textbf{Retrieval Outcomes}} 
      & \textbf{Invalid}
      & \multicolumn{3}{c}{\textbf{Accuracy@N}} 
      & \textbf{F1 Score} \\
      \cmidrule(lr){3-5} \cmidrule(lr){6-8} \cmidrule(lr){10-12}
      \textbf{Category}
        & \textbf{Videos}
        & \textbf{1} & \textbf{2} & \textbf{3}
        & \textbf{TN} & \textbf{FP} & \textbf{FN} & \textbf{(FP + TN)}
        & \textbf{@1} & \textbf{@2} & \textbf{@3}
        & \textbf{@1} \\
      \midrule
      Physics \& Collision & 36 & 28 & 31 & 32 & 9 & 7 & 2 & 8 (5 + 3) & 0.93 & 0.94 & 0.94 & 0.86 \\
      Animation \& VFX     & 23 & 11 & 16 & 16 & 13 & 19 & 2 & 21 (14 + 7) & 0.85 & 0.89 & 0.89 & 0.51 \\
      Rendering \& Texture & 32 & 22 & 27 & 29 & 8 & 13 & 0 & 16 (10 + 6) & 1.00 & 1.00 & 1.00 & 0.77 \\
      UI \& HUD            & 39 & 31 & 32 & 32 & 5 & 11 & 2 & 6 (6 + 0) & 0.94 & 0.94 & 0.94 & 0.83 \\
      Performance          & 38 & 20 & 20 & 20 & 11 & 6 & 13 & 2 (1 + 1) & 0.61 & 0.61 & 0.61 & 0.68 \\
      Lighting \& Shadow   & 49 & 40 & 44 & 44 & 1 & 5 & 0 & -- & 1.00 & 1.00 & 1.00 & 0.94 \\
      Game Crash \& Logic  & 30 & 17 & 20 & 22 & 17 & 9 & 2 & 4 (3 + 1) & 0.89 & 0.91 & 0.92 & 0.76 \\
      \midrule
      \textbf{Overall}     
        & \textbf{247} 
        & \textbf{169} & \textbf{190} & \textbf{195} 
        & \textbf{64} & \textbf{70} & \textbf{21} & \textbf{57 (39 + 18)}
        & \textbf{0.89} & \textbf{0.90} & \textbf{0.90} 
        & \textbf{0.79} \\
      \bottomrule
    \end{tabular}
  \end{threeparttable}
  }
\end{table*}

\begin{figure}
    \centering
    \includegraphics[width=0.9\linewidth]{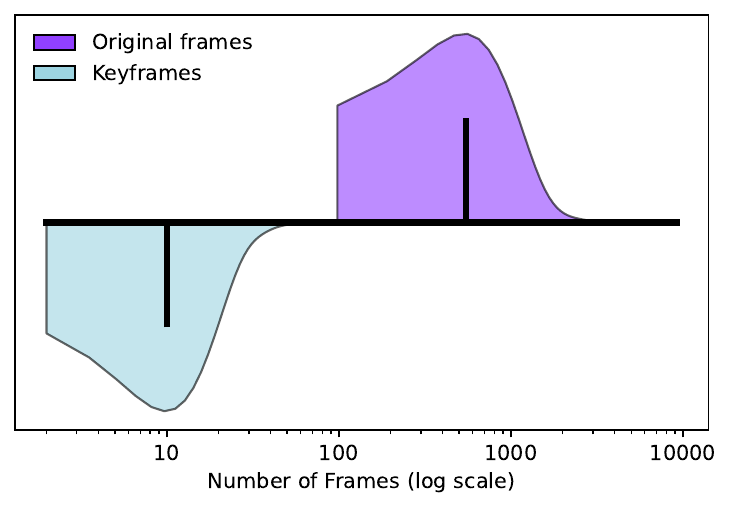}
    \caption{Total number of frames vs keyframes per video across 247 videos in log scale (bar indicates median value)}
    \label{fig:rq1_log_scale_frame}
\end{figure}

\section{Experimental setup}
\label{sec:exp_setup}

In this section, we first describe our experimental setup and then outline the performance metrics used in the evaluation.

\textbf{Used VLM/LLM models:} We used OpenAI’s GPT-4o~\cite{GPT4O} and GPT-4o-mini~\cite{GPT4O-mini}, accessed through API version ``2024-10-21" (at the time stable GA version)~\cite{Azure}, as well as the open source Mistral-7B model~\cite{mistral}. GPT-4o was selected for its strong performance in vision-language tasks, particularly for detecting visual glitches in game images~\cite{taesiri2025videogameqabenchevaluatingvisionlanguagemodels}.  GPT-4o-mini and Mistral-7B are used in the clustering stage of our study as discussed below.

\textbf{Data Collection:} We utilize real-world industrial game development data logged in JIRA, a widely used platform for bug tracking, issue tracking, and agile project management~\cite{Jira}. Specifically, our data consists of JIRA bug reports from a popular First-Person Shooter (FPS) game. Each report typically contains a textual bug description and one or more short gameplay videos (usually under one minute), which are submitted by developers or QA testers to demonstrate the bug. 

To evaluate retrieval accuracy and category-based performance in a consistent manner, we constructed a subset of 350 gameplay videos as follows:
\begin{itemize}
   \item We first summarized all JIRA bug reports from the past two years into preliminary keyword-based clusters using the Mistral-7B model. This initial step with an open source model was designed to reduce the computational cost of the GPT model due to the large number of reports.
    \item While Mistral-7B was effective for initial keyword-based clustering, it lacked the summarization quality needed to produce concise category names. We therefore used GPT-4o-mini to refine these clusters into seven high-level bug categories. The final category names are shown in Table~\ref{tab:category-definitions}.
    \item We then sampled JIRA reports from recent months that contained one or more video attachments, resulting in 2,880 reports. Each report was automatically assigned to one of the seven categories based on its textual description using GPT-4o-mini.
    \item From each of the seven categories, we sampled 50 video samples. One of the authors manually verified each sample to ensure it clearly aligned with the assigned category.
    \item We manually annotated the keyframes in the videos as bug or non-bug, if any. Note that several frames may represent the bug accurately. These keyframes will serve as ground truth frames in later evaluation.
\end{itemize}

These 350 videos are used in RQ2 and RQ3. 
For RQ1, a subset of 247 (manually confirmed to contain a clear visual bug) out of the 350 videos is used to evaluate keyframe extraction effectiveness, as discussed in Section~\ref{sec:rq_keyframes}.



\subsubsection{Performance evaluation:}
To evaluate the effectiveness of our bug frame retrieval pipeline, we consider two primary use cases: \textbf{Top-1 retrieval} and \textbf{Top-N retrieval}. These correspond to different levels of automation in game development workflows.
\begin{itemize}
    \item In the \textbf{Top-1 retrieval} use case, only the highest-ranked frame returned by the LLM is evaluated. This use case is aligned with scenarios where a fully automated system is desired, for example, automatically generating bug reports without human intervention. High precision in this setting is essential, as the top result alone must accurately represent the described bug.
    \item In the \textbf{Top-N retrieval} use case, we consider the first N-ranked frames. This reflects a use case in game development pipelines, where a human is still in the loop. For example, a QA engineer might inspect just the top 3 ranked keyframes to verify bugs mentioned in the report instead of watching a full gameplay video. In such workflows, the key objective is to ensure that at least one of the returned keyframes displays the described bug, reducing the time required to manually review entire videos.
\end{itemize}

To assess the correctness of results returned by the VLM, we compute the following metrics:

\begin{itemize}
    \item \textbf{Accuracy@N:} The fraction of videos for which a ground truth bug frame appears among the top $N$ retrieved frames. Formally, 
    \[
    \mathrm{Accuracy@}N = \frac{\mathrm{TP}@N}{\mathrm{TP}@N + \mathrm{FN}}
    \]

    \item \textbf{F1 Score@1:} Defined at Top-1 retrieval as:
    \[
    \mathrm{F1@1} = 2 \times \frac{\mathrm{TP}@1}{2 \cdot \mathrm{TP}@1 + \mathrm{FP} + \mathrm{FN}}
    \]
    where $\mathrm{TP}@1$ denotes true positives at top-1. 
    
    \textit{Note:} In our setup, false positives include those arising from invalid cases, and any retrieved frame that does not correspond to a valid bug frame, regardless of its rank. Therefore, $\mathrm{FP}@1 = \mathrm{FP}@2 = \mathrm{FP}@3 = \mathrm{FP}$, as well as $\mathrm{FN}$, are not position-specific.
\end{itemize}

We defined the components of the confusion matrix as follows:

\begin{itemize}
    \item \textbf{True Positive (TP):} The model successfully retrieved a bug frame.
    \item \textbf{True Negative (TN):} No visual bug is present in the video, and the model correctly returned no results.
    \item \textbf{False Positive (FP):} The model incorrectly identified a non-bug frame as a bug frame.
    \item \textbf{False Negative (FN):} A visual bug is present in the video, but the model failed to return any frame.
    \item \textbf{Invalid:} All keyframes extracted from the video did not include the bug moment. In such cases, the correct model behavior is to return no result (i.e. treated as TN if none were returned and FP otherwise).
\end{itemize}

We define the following metrics:
\begin{itemize}
    \item \textbf{Bug coverage:} The percentage of videos for which at least one extracted keyframe depicts the ground-truth bug.
    \item \textbf{Successful retrieval rate:} The proportion of bug-containing videos or reports where the pipeline successfully retrieves at least one frame that matches the ground-truth bug.
\end{itemize}

\section{Results}
\label{sec:results}
In this section, we answer our research questions.
\subsection{RQ 1: \RQframeselect}
\label{sec:rq_keyframes}
\textit{Motivation:} By manually analyzing the effectiveness of keyframe-based video segmentation, we assess whether extracting keyframes serves as an effective down-sampling technique for preserving bug-relevant content.

\textit{Approach:} We first curated a set of 247 of the studied 350 videos that contain visible visual bugs (bugs that can be easily spotted without ambiguity). Each video was manually verified by one of the authors to ensure it included at least one identifiable bug frame. This step guaranteed that the videos actually contained valid bug frames that could be extracted as keyframes. 
Second, for each video, we manually reviewed all keyframes produced by the extraction stage and recorded three metrics: 
(1)~the number of failures (defined as cases where no keyframe captured the bug), 
(2)~the average keyframe extraction rate (percentage of keyframes among all frames in a video) and 
(3)~the average number of keyframes per video that actually show the visual bug. 
To assess how video encoding configurations affect coverage, we conducted this analysis on the videos before and after re-encoding (as described in Section~\ref{sec:keyfream_extract}).

\textit{Findings:}
\textbf{After re-encoding, bug coverage reaches 98.79\% with a keyframe extraction rate of 1.90\%.} 
Given a median video length of 551 frames (Figure~\ref{fig:rq1_log_scale_frame}), a keyframe extraction rate of 1.90\% corresponds to approximately 10 keyframes per video, \emph{of which an average of 4.57 keyframes actually depict the bug}. Figure~\ref{fig:rq1_log_scale_frame} illustrates that the number of extracted keyframes is substantially lower than the total frame count per video. To quantify this reduction, we computed the Wilcoxon signed-rank test (a non-parametric test that examines whether the median paired difference \(W\) equals zero) and the Cliff's Delta ($\delta$) effect size between the distributions of total frames and extracted keyframes~\cite{cliff1993, Wilcoxon1992}. The two tests yielded  $W = 0$ and a large effect size (Cliff's $\delta = 0.99 $). These findings demonstrate that keyframe extraction is an effective down-sampling strategy.

\textbf{Extracting keyframes from unnormalized videos achieves only 65.99\% bug coverage at 1.10\% keyframe extraction rate.} 
When using each video’s original format, keyframes could not be extracted at all in 46 out of 247 videos, due to missing internal metadata or incompatible encoder settings. In the remaining 201 videos, 38 failed to include any keyframes that depicted the described bug. Even when excluding the 46 extraction failures, the remaining videos yield only 84.61\% coverage. On average, 1.10\% of frames were extracted per video (approximately 6 keyframes), and \emph{only 1.93 of those keyframes, on average, showed the bug}. These results suggest that lower bug coverage in the default setting is due to both (a) an insufficient number of keyframes extracted and (b) complete extraction failure in some videos. 
Original encoders may use long fixed keyframe intervals, low scene change sensitivity, or non-standard encoder formats that limit FFmpeg's ability to extract keyframes.
These findings highlight the importance of re-encoding, not just to increase keyframe density but to ensure keyframe extraction succeeds at all.

\vspace{5pt}
\noindent
\fbox{\parbox{\columnwidth}{
  \textit{Keyframes can accurately capture bug moments in gameplay videos. Re-encoding developer-submitted videos can significantly improve the keyframe extraction process.
}
}}

\subsection{RQ 2: \RQhowwell}
\textit{Motivation:} 
Understanding how accurately the model retrieves bug-relevant frames from a set of keyframes provides key insight into the practical utility of our pipeline. Specifically, it reflects how well the model can help locate the frame that matches the given bug description. Meanwhile, by evaluating the ability to retrieve bug frames in different categories of bugs, we can better understand its strengths and limitations and identify which types of visual bugs are most suitable for this type of automated analysis.

\textit{Approach:}  
In practice, developers often attach multiple videos to a single bug report. In fact, such cases account for \textbf{35\%} of our dataset. To account for this, we evaluated retrieval accuracy at two levels:

\begin{itemize}
  \item \textbf{Per-video accuracy:}  
     We evaluate the selected set of 350 videos as discussed in Section~\ref{sec:exp_setup} and assess performance under two retrieval settings: \textbf{Top-1} and \textbf{Top-N}.

  \item \textbf{Per-report accuracy:}  
    Not every video necessarily contains the reported visual bug, e.g. because developers may upload videos to demonstrate correct game behavior for comparison. To assess performance at the report level, we randomly selected 100 JIRA bug reports containing multiple video attachments and evaluated whether our pipeline could retrieve at least one relevant bug frame from any of the associated videos.

\end{itemize}

    We then conducted \textbf{per-category analysis:}  
    Each of the 50 samples from the 7 categories (Table~\ref{tab:category-definitions}) is sent to our pipeline to obtain ranked bug frame results.  For each sample, we assess the retrieval outcome against our manual annotation (ground truth keyframes containing visible bugs) and compute both the accuracy and F1 scores to quantify category-specific performance. We also analyze true positive, false positive, and false negative cases to identify limitations specific to each category.

\begin{figure}
    \centering
    \includegraphics[width=0.85\linewidth]{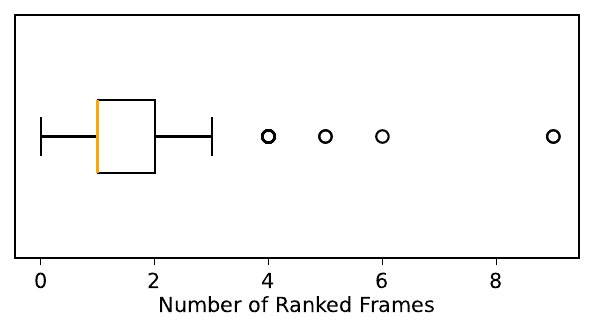}
    \caption{Distribution of number of frames in final ranked list}
    \label{fig:number_frame_rq3}
\end{figure}

\textit{Findings:}
\textbf{Per-video retrieval has an F1@1 score of 0.79 and an Accuracy@1 of 0.89.} Table~\ref{tab:frame-detection} shows that the model successfully retrieved a correct bug frame at the top-1 result for 169 out of 247 bug videos, yielding an Accuracy@1 of 0.89. When considering the top-3 results, performance improves to 195 videos, corresponding to an Accuracy@3 of 0.90. Across the evaluation set, we observe 70 false positives and 21 false negatives, resulting in an overall F1@1 score of 0.79.  The successful retrieval rate is 68.42\%. Figure~\ref{fig:number_frame_rq3} shows that the median number of frames selected per video is 1, corresponding to the top-1 results summarized in Table~\ref{tab:frame-detection}. This suggests that even when the model selects multiple candidate frames, the top-ranked frame is most often the correct one. In cases where multiple frames are selected, the matched bug frame always appears within the top 3 results.

\textbf{Per-report retrieval improves successful retrieval rate from 68.42\% to 82.14\%.}
Out of 100 sampled multi-video JIRA bug reports (337 videos total: 204 TP, 95 FP, 15 FN, and 23 TN), we identified 84 reports containing at least one video with a visible bug. Our pipeline successfully retrieved at least one correct bug frame in 69 of these 84 reports (82.14\%). While this does not affect the overall F1@1 score (which remains at 0.79), aggregating retrieval at the report level improves the chance of retrieving at least one valid bug frame per report.

\textbf{Lighting \& Shadow achieves the highest retrieval accuracy (F1@1 = 0.94)}, followed by Physics \& Collision and UI \& HUD. Table~\ref{tab:frame-detection} shows that the model achieves its highest performance on the Lighting \& Shadow category, retrieving bug frames more reliably than in other categories. Physics \& Collision and Camera \& UI also maintain high true-positive counts with low false-negative counts. Several common factors contribute to this higher performance trend:
\begin{itemize}
  \item \textbf{High visual contrast:} Bugs exhibit dramatic visual changes in the game, such as extremely dark scenes (Figure.~\ref{fig:lighting}), collisions with surrounding objects (Figure.~\ref{fig:clipping}), or placeholder images/text in UI elements that have colors that stand out from the rest of the environment (Figure.~\ref{fig:ui}).
  \item \textbf{Strong text–visual alignment:} Bug descriptions align well with visible content (e.g., ``character can clip underground” or ``placeholder text appears in the menu”), making it easier for the model to retrieve a correct frame.
\end{itemize}

Our analysis reveals several key limitations that hinder retrieval performance:
\begin{itemize}
 \item \textbf{Flickering events:} A major issue among rendering-related bugs is brief flickers of objects (e.g., a light source at a distance starts flickering). While these are easily noticeable during video playback, a single extracted frame often fails to capture the flicker.
\item \textbf{On-screen debug messages:} While videos in other categories include debug overlays as well, they’re not defining features of those categories. These messages particularly affect \textbf{Game Crash \& Logic} videos. Development logs and messages often dominate the visual focus and may redirect VLM to producing irrelevant retrieval.

\item \textbf{Abstract bugs:} Logic-related issues, such as a reward system not updating after an in-game action, often fall under \textbf{Game Crash \& Logic}. These bugs lack a clear visual signal and usually require reasoning across multiple frames to understand cause and effect.

\item \textbf{Post-processing effects mistaken as false positives:} Visual effects such as motion blur, bloom, and lens flares are prevalent in \textbf{Animation \& VFX} videos and usually trigger a new keyframe due to a major shift in visual content. These artistic effects are sometimes misinterpreted by the model as visual glitches, increasing false positive rates when analyzing isolated frames.

\item \textbf{Minimal visual anomalies:} While subtle artifacts can appear across categories, bugs in Animation \& VFX often involve minor animation glitches that are difficult to observe (Figure~\ref{fig:animation}).

\item \textbf{Frame-rate drops:} Most \textbf{Performance} bugs involve degraded frame rates during gameplay. Although developer-submitted videos often include an on-screen FPS indicator, the expected frame rate is rarely specified, and these drops often do not trigger new keyframes due to minimal visual changes. As a result, the model struggles to identify frame-rate issues, especially in modern games where high frame rates are standard. For instance, a drop to 90 FPS may indicate poor performance if the game is expected to run at 120 FPS, just as 40 FPS would be problematic in a game targeting 60 FPS. Without explicit context, the model cannot reliably infer whether the observed frame rate reflects a bug.
 \end{itemize}

\vspace{5pt}
\noindent
\fbox{\parbox{\columnwidth}{
  \textit{Our pipeline achieves an Accuracy@1 of 0.89 and an F1@1 score of 0.79 at the per-video level and retrieves at least one bug frame in 82.14\% of multi-video reports. Retrieval is highest in \textbf{Lighting \& Shadow}, \textbf{Physics \& Collision}, and \textbf{UI \& HUD} categories, and the weakest in \textbf{Performance} and \textbf{Animation \& VFX}. These results demonstrate the practical effectiveness of VLM-based retrieval of bug frames in gameplay videos and the potential to considerably reduce video viewing time.}
}}

\begin{table}[htbp]
  \centering
  \small             
  \caption{Retrieval performance across three identical runs}
  \label{tab:rq4_stable}

  \begin{threeparttable}
    \begin{tabular}{c c c c c c c c}
      \toprule
      &
      \textbf{TP@1} &
      \textbf{TP@2} &
      \textbf{TP@3} &
      \textbf{FP} &
      \textbf{TN} &
      \textbf{FN} &
      \textbf{F1@1}\\
      \midrule
      1 & 169 & 190 & 195 & 70 & 64 & 21 & 0.79 \\
      2 & 180 & 194 & 194 & 72 & 62 & 22 & 0.79 \\
      3 & 179 & 192 & 192 & 76 & 58 & 24 & 0.78\\
       \midrule
       \textbf{Total} & 528 & 576 & 581 & 218 & 184 & 67 \\
      \bottomrule
    \end{tabular}

  \end{threeparttable}
\end{table}

\subsection{RQ 3: \RQconfidence}
\textit{Motivation:} A known limitation of large language models is their non-deterministic behavior: repeated queries to the same model can yield different outputs, even when the input remains unchanged. This variability stems from the probabilistic nature of their design and sampling mechanisms~\cite{peeperkorn2024temperaturecreativityparameterlarge, lovering2025languagemodelprobabilitiescalibrated, vidler2025playinggameslargelanguage, klishevich2025measuringdeterminismlargelanguage}. While some degree of output variation is expected, it raises concerns about the reliability and reproducibility of our pipeline. Therefore, we assess how consistent the retrieval results are across multiple runs of the same prompt and setup.

\textit{Approach:}
We re-run the same set of 350 videos two additional times, resulting in a total of three runs under identical experimental conditions. For each run, we assess the retrieval results at \textbf{Top-1} using the same evaluation metrics as before. This allows us to measure the consistency of the pipeline's outputs across repeated executions. To quantify the likelihood of successful retrieval across multiple runs, we compute \textbf{pass@k}, which denotes the probability that a correct bug frame is retrieved in at least one of the k runs.

\textit{Findings:} As shown in Table~\ref{tab:rq4_stable}, the F1@1 score varies minimally across the three runs, with a median of 0.79. However, despite this stability in the overall metrics, there is some inconsistency across runs for many of the videos.
For example, a bug frame might be extracted correctly in run 1 but incorrectly in run 2.

\begin{itemize}

\item{\textbf{\# of videos correct in 3/3 runs:}} A total of 183 (140 TP and 43 TN) videos had correct retrieval outcomes in all three runs.

\item{\textbf{\# of videos correct in 2/3 runs:}} 48 (42 TP and 6 TN) videos had correct retrieval outcomes in two of the three runs.

\item{\textbf{\# of videos correct in 1/3 runs:}} 25 (24 TP and 1 TN) videos had correct retrieval outcomes in only one of the runs.

\item{\textbf{\# of videos correct in 0/3 runs:}} 94 videos including FP and FN. Since only TP@1 is considered, videos where the correct bug frame was retrieved at rank 2 or 3 are considered as incorrect in this analysis.
\end{itemize}

\vspace{4pt}
\noindent\textbf{Single-run accuracy (\textit{pass@1}):}  
The \textit{pass@1} metric captures the probability that the pipeline retrieves a correct result in a single run@1. It is defined as:

\[
\text{pass@1} = \frac{\text{TP}@1 + \text{TN}\text{ across 3 runs}}{\text{\#Videos across 3 runs}} = \frac{528 + 184}{350\times3} \approx 0.68.
\]

\vspace{4pt}
\noindent\textbf{Three-run accuracy (pass@3):} The probability that \emph{at least one} of the three
runs@3 succeeds:

\[
\text{pass@3} = 1 - (1 - \text{pass@1})^3 \approx0.98
\]

This means that under one run, there is a 68\% chance that the top-1 retrieved frame is a correct result. This probability increases to 98\% under 3 runs.

\noindent\textbf{Expected value\& variation of correctness:}  
We define the variance of the random variable \( X \in \left\{0, \frac{1}{3}, \frac{2}{3}, 1\right\} \) and the expected number of successful retrievals per video as:
\[
\mathbb{E}(X) = \frac{
  94 \cdot 0 +
  25 \cdot \frac{1}{3} +
  48 \cdot \frac{2}{3} +
  183 \cdot 1.0
}{350}
\approx 0.63
\]
To quantify the variability in correctness:
\[
\text{Var}(X) = \mathbb{E}(X^2) - \mathbb{E}(X)^2
\]
\[
\mathbb{E}(X^2) = 0^2 \cdot \frac{94}{350}
+ \left(\frac{1}{3}\right)^2 \cdot \frac{25}{350}
+ \left(\frac{2}{3}\right)^2 \cdot \frac{48}{350}
+ 1^2 \cdot \frac{183}{350}
\approx 0.59
\]
\[
\text{Var}(X) = 0.59 - (0.63)^2 \approx 0.19
\]
\[
\text{SD}(X) = \sqrt{\text{Var}(X)} \approx 0.44
\]

The expected value $\mathbb{E}(X) \approx 0.63$ means that, on average, each video yields correct retrievals in roughly 1.89 ($0.63\times3$) out of 3 runs. The variance $\text{Var}(X) \approx 0.19$ and corresponding standard deviation $0.44$ indicate considerable variability in correctness, and that using a single run per video may lead to unreliable results. 

\subsubsection{Variance Mitigation Recommendation}

To mitigate retrieval instability across runs, we recommend executing the pipeline multiple times (e.g., three runs per video) and applying a simple majority vote over the retrieved frames. This strategy helps improve consistency without requiring access to ground truth:

\begin{itemize}
  \item If the same bug frame or no result appears in at least two of the three runs, treat the result as reliable.
  \item Otherwise, flag the video as \emph{ambiguous} and consider reviewing it manually.
\end{itemize}

This approach filters out inconsistent results. While it introduces some computational overhead, it improves practical confidence in retrieval quality. In our evaluation, applying majority voting would raise the number of stable results from 183 to 215 out of 350 videos.

\vspace{5pt}
\noindent
\fbox{\parbox{\columnwidth}{
  \textit{Retrieval performance is stable overall (median F1@1 = 0.79), but per-video results vary.}  
  A majority vote over 3 runs helps improve consistency by confirming stable results and flagging ambiguous ones.

}}

\section{Discussion}
\label{sec:discussion}
\textbf{Comparison with prior work on the topic:} 
According to the recent benchmark study \textit{VideoGameQA-Bench}~\cite{taesiri2025videogameqabenchevaluatingvisionlanguagemodels} by Taesiri et al., GPT-4o is the top-tier model for image-based game glitch detection, achieving an accuracy of 82.9\% and an F1 score of 0.83 on the benchmark dataset.  However, this dataset is constructed from publicly available data, which predominantly features clear and easily recognizable glitches. In contrast, real-world QA processes often involve subtle visual issues that may be difficult to spot during normal gameplay. To evaluate GPT-4o's performance under such industrial conditions, we randomly sampled 100 images with a similar distribution (50 normal gameplay images and 50 known-bug images). Each image was processed by GPT-4o using the game glitch detection prompt defined in \textit{VideoGameQA-Bench}. When asked to classify whether an image contained a glitch, GPT-4o only achieved 46\% accuracy and an F1 score of 0.52 on our data. These results indicate that, in a development setting, pure image-based detection struggles with the complexity of our data and underscores the need to provide additional input context. In contrast, our retrieval pipeline leverages both visual (frame) and textual (bug description) inputs, achieving Accuracy@1 = 89\% and F1@1 = 0.79.

\textbf{Limitations of our approach:}
Based on our results, we recommend applying automated bug frame retrieval primarily to categories with strong visual cues, such as Lighting \& Shadow, Rendering \& Texture, and UI \& HUD.  In contrast, categories like Animation \& VFX or Performance are inherently temporal in nature and may require fundamentally different approaches, such as multi-frame analysis or video-level modeling. Future work should explore incorporating temporal context into the pipeline to handle such cases, as static frame retrieval alone is insufficient.

\textbf{Practical Impact:}
Our approach offers an efficient and scalable solution for reviewing gameplay videos. Given a median video length of 551 frames (approximately 9 seconds), we achieve a top-1 accuracy of 0.89, in nearly 9 out of 10 cases, a single retrieved frame correctly captures the visual bug. This reduces manual review effort from hundreds of frames to just one. Although the median video durations seem short, it is important to realize that it is often necessary to watch a video several times before spotting the bug. Hence, the actual benefits are larger.  Furthermore, the entire set of experiments was conducted for under \$100, demonstrating that the method is not only accurate but also cost-effective.

\section{Threats to Validity}
\label{Sec:threats}

\textit{\textbf{Internal validity:}}
A potential threat to the internal validity of our results is the used encoding settings. We used the default FFmpeg settings to re-encode the videos. As these settings achieved a 98.79\% bug coverage rate in our preliminary study, we did not fine tune these settings any further. However, there could be settings that result in an even higher bug coverage rate and a lower frame extraction rate. Future studies should further investigate the impact of encoding settings on our approach.

Another threat arises from the batch size used during VLM analysis. Due to API limitations, we chose to process frames in batches for longer videos. If a bug-related sequence is split across batch boundaries or if the relevant context spans multiple batches, the VLM may lack sufficient visual continuity to retrieve the correct frame. Despite its negligible impact on our studied videos, its effect on long video performance should be investigated by future studies.

Additional threats arise from components within our own pipeline. Specifically, we use GPT-4o to generate summaries of bug descriptions and assume its correctness. Although its outputs achieved 97\% agreement with human judgment in a 100-sample evaluation, this sample may not fully represent the diversity of bug types encountered in the data. As a result, the reliability of these summaries remains uncertain during full pipeline execution.

Furthermore, the prompting strategies used in the pipeline are straightforward, relying primarily on zero-shot or minimally guided prompts without task-specific fine-tuning. Although this design promotes ease of integration, it may limit performance in more ambiguous scenarios. Changes in prompt phrasing, structure, or context can influence the interpretation and ranking behavior of the VLM and may produce different results.

Finally, manual evaluation of VLM‐generated ranked frames introduces potential bias. Determining whether a given frame accurately matches its bug description is sometimes subjective and ambiguous, particularly in difficult and subtle cases. These human judgments may inadvertently overestimate or underestimate the true performance of the model.

\textit{\textbf{External validity:}} A threat to the external validity is that the studied bug reports were all obtained from a single FPS (First Person Shooter) game. While the game features a variety of visual bug types and development workflows, the domain-specific nature of its mechanics, UI design, and bug reporting structure may not fully represent games in other genres.
For example, the presentation of animation bugs, UI glitches, or logic errors may vary significantly between game types and game engines. Future studies should further investigate our approach for other types of games. 


Another factor affecting external validity is the dependency on a specific version of the language model. Our pipeline is built entirely on the GPT-4o model, which is subject to updates, deprecation, or behavioral changes over time. The same prompts and inputs may yield different outputs in future versions or different models due to changes in model architecture, training data, backend configuration, etc. This model-specific dependency raises concerns about the long-term robustness of the approach, especially in production environments where model availability may change over time, causing performance variations.

\section{Conclusion}
\label{sec:conclu}
In this study, we introduced an automated pipeline combining keyframe extraction and GPT-4o-based frame analysis to effectively retrieve bug frames from gameplay videos. Our pipeline demonstrates strong retrieval performance on real-world JIRA bug data from a popular FPS game, achieving an overall F1@1 score of 0.79 and an Accuracy@1 of 0.89. The method showed particularly strong results in retrieving bug frames related to Lighting \& Shadow, Physics \& Collision, and UI \& HUD. We identified key limitations, particularly in Animation \& VFX bugs, where the pipeline struggled mostly due to insufficient context captured by individual frames. Future work could enhance performance by incorporating temporal context analysis, fine-tuning prompting strategies, and evaluating our pipeline across diverse game genres and more extensive datasets. Overall, our approach presents an important, practical advancement in automated game quality assurance, significantly reducing manual effort and providing robust support for visual bug retrieval in real-world game development scenarios.

\bibliographystyle{ACM-Reference-Format}
\bibliography{references}


\begin{thebibliography}{53}


\ifx \showCODEN    \undefined \def \showCODEN     #1{\unskip}     \fi
\ifx \showDOI      \undefined \def \showDOI       #1{#1}\fi
\ifx \showISBNx    \undefined \def \showISBNx     #1{\unskip}     \fi
\ifx \showISBNxiii \undefined \def \showISBNxiii  #1{\unskip}     \fi
\ifx \showISSN     \undefined \def \showISSN      #1{\unskip}     \fi
\ifx \showLCCN     \undefined \def \showLCCN      #1{\unskip}     \fi
\ifx \shownote     \undefined \def \shownote      #1{#1}          \fi
\ifx \showarticletitle \undefined \def \showarticletitle #1{#1}   \fi
\ifx \showURL      \undefined \def \showURL       {\relax}        \fi
\providecommand\bibfield[2]{#2}
\providecommand\bibinfo[2]{#2}
\providecommand\natexlab[1]{#1}
\providecommand\showeprint[2][]{arXiv:#2}

\bibitem[FFm(2025)]%
        {FFmpeg}
 \bibinfo{year}{2025}\natexlab{}.
\newblock \bibinfo{booktitle}{\emph{FFmpeg}}.
\newblock
\urldef\tempurl%
\url{https://ffmpeg.org/}
\showURL{%
\tempurl}


\bibitem[Anthropic(2024)]%
        {Claude}
\bibfield{author}{\bibinfo{person}{Anthropic}.} \bibinfo{year}{2024}\natexlab{}.
\newblock \bibinfo{booktitle}{\emph{Introducing the next generation of Claude}}.
\newblock
\urldef\tempurl%
\url{https://www.anthropic.com/news/claude-3-family}
\showURL{%
\tempurl}


\bibitem[Ariyurek et~al\mbox{.}(2021)]%
        {auto_game_test}
\bibfield{author}{\bibinfo{person}{Sinan Ariyurek}, \bibinfo{person}{Aysu Betin-Can}, {and} \bibinfo{person}{Elif Surer}.} \bibinfo{year}{2021}\natexlab{}.
\newblock \showarticletitle{Automated Video Game Testing Using Synthetic and Humanlike Agents}.
\newblock \bibinfo{journal}{\emph{IEEE Transactions on Games}} \bibinfo{volume}{13}, \bibinfo{number}{1} (\bibinfo{year}{2021}), \bibinfo{pages}{50--67}.
\newblock
\urldef\tempurl%
\url{https://doi.org/10.1109/TG.2019.2947597}
\showDOI{\tempurl}


\bibitem[Atlassian(2025)]%
        {Jira}
\bibfield{author}{\bibinfo{person}{Atlassian}.} \bibinfo{year}{2025}\natexlab{}.
\newblock \bibinfo{booktitle}{\emph{Jira}}.
\newblock
\urldef\tempurl%
\url{https://www.atlassian.com/software/jira}
\showURL{%
\tempurl}


\bibitem[Azizi and Zaman(2023)]%
        {azizi2023automaticbugdetectiongames}
\bibfield{author}{\bibinfo{person}{Elham Azizi} {and} \bibinfo{person}{Loutfouz Zaman}.} \bibinfo{year}{2023}\natexlab{}.
\newblock \showarticletitle{Automatic Bug Detection in Games using LSTM Networks}. In \bibinfo{booktitle}{\emph{2023 IEEE Conference on Games (CoG)}}. \bibinfo{pages}{1--4}.
\newblock
\urldef\tempurl%
\url{https://doi.org/10.1109/CoG57401.2023.10333253}
\showDOI{\tempurl}


\bibitem[Bettenburg et~al\mbox{.}(2008)]%
        {whatmakesgoodbugreport}
\bibfield{author}{\bibinfo{person}{Nicolas Bettenburg}, \bibinfo{person}{Sascha Just}, \bibinfo{person}{Adrian Schr\"{o}ter}, \bibinfo{person}{Cathrin Weiss}, \bibinfo{person}{Rahul Premraj}, {and} \bibinfo{person}{Thomas Zimmermann}.} \bibinfo{year}{2008}\natexlab{}.
\newblock \showarticletitle{What makes a good bug report?}. In \bibinfo{booktitle}{\emph{Proceedings of the 16th ACM SIGSOFT International Symposium on Foundations of Software Engineering}} (Atlanta, Georgia) \emph{(\bibinfo{series}{SIGSOFT '08/FSE-16})}. \bibinfo{publisher}{Association for Computing Machinery}, \bibinfo{address}{New York, NY, USA}, \bibinfo{pages}{308–318}.
\newblock
\showISBNx{9781595939951}
\urldef\tempurl%
\url{https://doi.org/10.1145/1453101.1453146}
\showDOI{\tempurl}


\bibitem[Campanella et~al\mbox{.}(2024)]%
        {Replication_Game_Debugging}
\bibfield{author}{\bibinfo{person}{Stefano Campanella}, \bibinfo{person}{Emanuela Guglielmi}, \bibinfo{person}{Rocco Oliveto}, \bibinfo{person}{Gabriele Bavota}, {and} \bibinfo{person}{Simone Scalabrino}.} \bibinfo{year}{2024}\natexlab{}.
\newblock \showarticletitle{Towards the Automatic Replication of Gameplays to Support Game Debugging}. In \bibinfo{booktitle}{\emph{Proceedings of the 1st ACM International Workshop on Foundations of Applied Software Engineering for Games}} (Porto de Galinhas, Brazil) \emph{(\bibinfo{series}{FaSE4Games 2024})}. \bibinfo{publisher}{Association for Computing Machinery}, \bibinfo{address}{New York, NY, USA}, \bibinfo{pages}{1–6}.
\newblock
\showISBNx{9798400706745}
\urldef\tempurl%
\url{https://doi.org/10.1145/3663532.3664465}
\showDOI{\tempurl}


\bibitem[Cliff(1993)]%
        {cliff1993}
\bibfield{author}{\bibinfo{person}{Norman Cliff}.} \bibinfo{year}{1993}\natexlab{}.
\newblock \bibinfo{booktitle}{\emph{Dominance Statistics: Ordinal Analyses to Answer Ordinal Questions}}. Vol.~\bibinfo{volume}{114}.
\newblock \bibinfo{pages}{494--509}.
\newblock


\bibitem[Cooper et~al\mbox{.}(2021)]%
        {cooper2021takestangocombiningvisual}
\bibfield{author}{\bibinfo{person}{Nathan Cooper}, \bibinfo{person}{Carlos Bernal-C\'{a}rdenas}, \bibinfo{person}{Oscar Chaparro}, \bibinfo{person}{Kevin Moran}, {and} \bibinfo{person}{Denys Poshyvanyk}.} \bibinfo{year}{2021}\natexlab{}.
\newblock \showarticletitle{It Takes Two to Tango: Combining Visual and Textual Information for Detecting Duplicate Video-Based Bug Reports}. In \bibinfo{booktitle}{\emph{Proceedings of the 43rd International Conference on Software Engineering}} (Madrid, Spain) \emph{(\bibinfo{series}{ICSE '21})}. \bibinfo{publisher}{IEEE Press}, \bibinfo{pages}{957–969}.
\newblock
\showISBNx{9781450390859}
\urldef\tempurl%
\url{https://doi.org/10.1109/ICSE43902.2021.00091}
\showDOI{\tempurl}


\bibitem[Dong et~al\mbox{.}(2024)]%
        {dong2024llmpersonalizedjudge}
\bibfield{author}{\bibinfo{person}{Yijiang~River Dong}, \bibinfo{person}{Tiancheng Hu}, {and} \bibinfo{person}{Nigel Collier}.} \bibinfo{year}{2024}\natexlab{}.
\newblock \bibinfo{title}{Can LLM be a Personalized Judge?}
\newblock
\newblock
\showeprint[arxiv]{2406.11657}~[cs.CL]
\urldef\tempurl%
\url{https://arxiv.org/abs/2406.11657}
\showURL{%
\tempurl}


\bibitem[EA(2025)]%
        {EA_forum}
\bibfield{author}{\bibinfo{person}{EA}.} \bibinfo{year}{2025}\natexlab{}.
\newblock \bibinfo{booktitle}{\emph{EA Forum.}}
\newblock
\urldef\tempurl%
\url{https://forums.ea.com/}
\showURL{%
\tempurl}


\bibitem[Feng and Chen(2022)]%
        {feng2022gifdroidautomatedreplayvisual}
\bibfield{author}{\bibinfo{person}{Sidong Feng} {and} \bibinfo{person}{Chunyang Chen}.} \bibinfo{year}{2022}\natexlab{}.
\newblock \showarticletitle{GIFdroid: automated replay of visual bug reports for Android apps}. In \bibinfo{booktitle}{\emph{Proceedings of the 44th International Conference on Software Engineering}} (Pittsburgh, Pennsylvania) \emph{(\bibinfo{series}{ICSE '22})}. \bibinfo{publisher}{Association for Computing Machinery}, \bibinfo{address}{New York, NY, USA}, \bibinfo{pages}{1045–1057}.
\newblock
\showISBNx{9781450392211}
\urldef\tempurl%
\url{https://doi.org/10.1145/3510003.3510048}
\showDOI{\tempurl}


\bibitem[Gallotta et~al\mbox{.}(2024)]%
        {Roadmap_LLM_game}
\bibfield{author}{\bibinfo{person}{Roberto Gallotta}, \bibinfo{person}{Graham Todd}, \bibinfo{person}{Marvin Zammit}, \bibinfo{person}{Sam Earle}, \bibinfo{person}{Antonios Liapis}, \bibinfo{person}{Julian Togelius}, {and} \bibinfo{person}{Georgios~N. Yannakakis}.} \bibinfo{year}{2024}\natexlab{}.
\newblock \showarticletitle{Large Language Models and Games: A Survey and Roadmap}.
\newblock \bibinfo{journal}{\emph{IEEE Transactions on Games}} (\bibinfo{year}{2024}), \bibinfo{pages}{1--18}.
\newblock
\urldef\tempurl%
\url{https://doi.org/10.1109/TG.2024.3461510}
\showDOI{\tempurl}


\bibitem[Google(2024)]%
        {Gemini2.0}
\bibfield{author}{\bibinfo{person}{Google}.} \bibinfo{year}{2024}\natexlab{}.
\newblock \bibinfo{booktitle}{\emph{Introducing Gemini 2.0: our new AI model for the agentic era}}.
\newblock
\urldef\tempurl%
\url{https://blog.google/technology/google-deepmind/google-gemini-ai-update-december-2024/}
\showURL{%
\tempurl}


\bibitem[Gu et~al\mbox{.}(2025)]%
        {gu2025surveyllmasajudge}
\bibfield{author}{\bibinfo{person}{Jiawei Gu}, \bibinfo{person}{Xuhui Jiang}, \bibinfo{person}{Zhichao Shi}, \bibinfo{person}{Hexiang Tan}, \bibinfo{person}{Xuehao Zhai}, \bibinfo{person}{Chengjin Xu}, \bibinfo{person}{Wei Li}, \bibinfo{person}{Yinghan Shen}, \bibinfo{person}{Shengjie Ma}, \bibinfo{person}{Honghao Liu}, \bibinfo{person}{Saizhuo Wang}, \bibinfo{person}{Kun Zhang}, \bibinfo{person}{Yuanzhuo Wang}, \bibinfo{person}{Wen Gao}, \bibinfo{person}{Lionel Ni}, {and} \bibinfo{person}{Jian Guo}.} \bibinfo{year}{2025}\natexlab{}.
\newblock \bibinfo{title}{A Survey on LLM-as-a-Judge}.
\newblock
\newblock
\showeprint[arxiv]{2411.15594}~[cs.CL]
\urldef\tempurl%
\url{https://arxiv.org/abs/2411.15594}
\showURL{%
\tempurl}


\bibitem[Guglielmi et~al\mbox{.}(2025)]%
        {Stuttering}
\bibfield{author}{\bibinfo{person}{Emanuela Guglielmi}, \bibinfo{person}{Gabriele Bavota}, \bibinfo{person}{Rocco Oliveto}, {and} \bibinfo{person}{Simone Scalabrino}.} \bibinfo{year}{2025}\natexlab{}.
\newblock \showarticletitle{Automatic Identification of Game Stuttering via Gameplay Videos Analysis}.
\newblock \bibinfo{journal}{\emph{ACM Trans. Softw. Eng. Methodol.}} \bibinfo{volume}{34}, \bibinfo{number}{2}, Article \bibinfo{articleno}{38} (\bibinfo{date}{Jan.} \bibinfo{year}{2025}), \bibinfo{numpages}{29}~pages.
\newblock
\showISSN{1049-331X}
\urldef\tempurl%
\url{https://doi.org/10.1145/3695992}
\showDOI{\tempurl}


\bibitem[Guglielmi et~al\mbox{.}(2022)]%
        {guglielmi2022usinggameplayvideosdetecting}
\bibfield{author}{\bibinfo{person}{Emanuela Guglielmi}, \bibinfo{person}{Simone Scalabrino}, \bibinfo{person}{Gabriele Bavota}, {and} \bibinfo{person}{Rocco Oliveto}.} \bibinfo{year}{2022}\natexlab{}.
\newblock \bibinfo{title}{Towards Using Gameplay Videos for Detecting Issues in Video Games}.
\newblock
\newblock
\showeprint[arxiv]{2204.04182}~[cs.SE]
\urldef\tempurl%
\url{https://arxiv.org/abs/2204.04182}
\showURL{%
\tempurl}


\bibitem[Guglielmi et~al\mbox{.}(2023)]%
        {guglielmi2023usinggameplayvideosdetecting}
\bibfield{author}{\bibinfo{person}{Emanuela Guglielmi}, \bibinfo{person}{Simone Scalabrino}, \bibinfo{person}{Gabriele Bavota}, {and} \bibinfo{person}{Rocco Oliveto}.} \bibinfo{year}{2023}\natexlab{}.
\newblock \showarticletitle{Using gameplay videos for detecting issues in video games}.
\newblock \bibinfo{journal}{\emph{Empirical Softw. Engg.}} \bibinfo{volume}{28}, \bibinfo{number}{6} (\bibinfo{date}{Oct.} \bibinfo{year}{2023}), \bibinfo{numpages}{32}~pages.
\newblock
\showISSN{1382-3256}
\urldef\tempurl%
\url{https://doi.org/10.1007/s10664-023-10365-0}
\showDOI{\tempurl}


\bibitem[Hu et~al\mbox{.}(2025)]%
        {hu2025surveylargelanguagemodelbased}
\bibfield{author}{\bibinfo{person}{Sihao Hu}, \bibinfo{person}{Tiansheng Huang}, \bibinfo{person}{Gaowen Liu}, \bibinfo{person}{Ramana~Rao Kompella}, \bibinfo{person}{Fatih Ilhan}, \bibinfo{person}{Selim~Furkan Tekin}, \bibinfo{person}{Yichang Xu}, \bibinfo{person}{Zachary Yahn}, {and} \bibinfo{person}{Ling Liu}.} \bibinfo{year}{2025}\natexlab{}.
\newblock \bibinfo{title}{A Survey on Large Language Model-Based Game Agents}.
\newblock
\newblock
\showeprint[arxiv]{2404.02039}~[cs.AI]
\urldef\tempurl%
\url{https://arxiv.org/abs/2404.02039}
\showURL{%
\tempurl}


\bibitem[Kaur et~al\mbox{.}(2024)]%
        {effective_keyframe_extraction}
\bibfield{author}{\bibinfo{person}{Sumandeep Kaur}, \bibinfo{person}{Lakhwinder Kaur}, {and} \bibinfo{person}{Madan Lal}.} \bibinfo{year}{2024}\natexlab{}.
\newblock \showarticletitle{An effective Key Frame Extraction technique based on Feature Fusion and Fuzzy-C means clustering with Artificial Hummingbird}.
\newblock \bibinfo{journal}{\emph{Scientific Reports}}  \bibinfo{volume}{14} (\bibinfo{date}{11} \bibinfo{year}{2024}).
\newblock
\urldef\tempurl%
\url{https://doi.org/10.1038/s41598-024-75923-y}
\showDOI{\tempurl}


\bibitem[Klishevich et~al\mbox{.}(2025)]%
        {klishevich2025measuringdeterminismlargelanguage}
\bibfield{author}{\bibinfo{person}{Eugene Klishevich}, \bibinfo{person}{Yegor Denisov-Blanch}, \bibinfo{person}{Simon Obstbaum}, \bibinfo{person}{Igor Ciobanu}, {and} \bibinfo{person}{Michal Kosinski}.} \bibinfo{year}{2025}\natexlab{}.
\newblock \bibinfo{title}{Measuring Determinism in Large Language Models for Software Code Review}.
\newblock
\newblock
\showeprint[arxiv]{2502.20747}~[cs.SE]
\urldef\tempurl%
\url{https://arxiv.org/abs/2502.20747}
\showURL{%
\tempurl}


\bibitem[Korolkov(2025)]%
        {korolkov2025scenedetectionpolicieskeyframe}
\bibfield{author}{\bibinfo{person}{Vasilii Korolkov}.} \bibinfo{year}{2025}\natexlab{}.
\newblock \bibinfo{title}{Scene Detection Policies and Keyframe Extraction Strategies for Large-Scale Video Analysis}.
\newblock
\newblock
\showeprint[arxiv]{2506.00667}~[cs.CV]
\urldef\tempurl%
\url{https://arxiv.org/abs/2506.00667}
\showURL{%
\tempurl}


\bibitem[Li et~al\mbox{.}(2024b)]%
        {li2024llava}
\bibfield{author}{\bibinfo{person}{Bo Li}, \bibinfo{person}{Yuanhan Zhang}, \bibinfo{person}{Dong Guo}, \bibinfo{person}{Renrui Zhang}, \bibinfo{person}{Feng Li}, \bibinfo{person}{Hao Zhang}, \bibinfo{person}{Kaichen Zhang}, \bibinfo{person}{Yanwei Li}, \bibinfo{person}{Ziwei Liu}, {and} \bibinfo{person}{Chunyuan Li}.} \bibinfo{year}{2024}\natexlab{b}.
\newblock \showarticletitle{LLaVA-OneVision: Easy Visual Task Transfer}.
\newblock \bibinfo{journal}{\emph{arXiv preprint arXiv:2408.03326}} (\bibinfo{year}{2024}).
\newblock


\bibitem[Li et~al\mbox{.}(2024a)]%
        {li2024llmsasjudgescomprehensivesurveyllmbased}
\bibfield{author}{\bibinfo{person}{Haitao Li}, \bibinfo{person}{Qian Dong}, \bibinfo{person}{Junjie Chen}, \bibinfo{person}{Huixue Su}, \bibinfo{person}{Yujia Zhou}, \bibinfo{person}{Qingyao Ai}, \bibinfo{person}{Ziyi Ye}, {and} \bibinfo{person}{Yiqun Liu}.} \bibinfo{year}{2024}\natexlab{a}.
\newblock \bibinfo{title}{LLMs-as-Judges: A Comprehensive Survey on LLM-based Evaluation Methods}.
\newblock
\newblock
\showeprint[arxiv]{2412.05579}~[cs.CL]
\urldef\tempurl%
\url{https://arxiv.org/abs/2412.05579}
\showURL{%
\tempurl}


\bibitem[Lin et~al\mbox{.}(2019)]%
        {Dayi_game_bug_video}
\bibfield{author}{\bibinfo{person}{Dayi Lin}, \bibinfo{person}{Cor-Paul Bezemer}, {and} \bibinfo{person}{Ahmed~E. Hassan}.} \bibinfo{year}{2019}\natexlab{}.
\newblock \showarticletitle{Identifying gameplay videos that exhibit bugs in computer games}.
\newblock \bibinfo{journal}{\emph{Empirical Software Engineering}} (\bibinfo{date}{12} \bibinfo{year}{2019}).
\newblock
\urldef\tempurl%
\url{https://doi.org/10.1007/s10664-019-09733-6}
\showDOI{\tempurl}


\bibitem[Lovering et~al\mbox{.}(2025)]%
        {lovering2025languagemodelprobabilitiescalibrated}
\bibfield{author}{\bibinfo{person}{Charles Lovering}, \bibinfo{person}{Michael Krumdick}, \bibinfo{person}{Viet~Dac Lai}, \bibinfo{person}{Seth Ebner}, \bibinfo{person}{Nilesh Kumar}, \bibinfo{person}{Varshini Reddy}, \bibinfo{person}{Rik Koncel-Kedziorski}, {and} \bibinfo{person}{Chris Tanner}.} \bibinfo{year}{2025}\natexlab{}.
\newblock \bibinfo{title}{Language Model Probabilities are Not Calibrated in Numeric Contexts}.
\newblock
\newblock
\showeprint[arxiv]{2410.16007}~[cs.AI]
\urldef\tempurl%
\url{https://arxiv.org/abs/2410.16007}
\showURL{%
\tempurl}


\bibitem[Ma et~al\mbox{.}(2024)]%
        {ma2024largelanguagemodelsplay}
\bibfield{author}{\bibinfo{person}{Weiyu Ma}, \bibinfo{person}{Qirui Mi}, \bibinfo{person}{Yongcheng Zeng}, \bibinfo{person}{Xue Yan}, \bibinfo{person}{Yuqiao Wu}, \bibinfo{person}{Runji Lin}, \bibinfo{person}{Haifeng Zhang}, {and} \bibinfo{person}{Jun Wang}.} \bibinfo{year}{2024}\natexlab{}.
\newblock \showarticletitle{Large Language Models Play StarCraft II:Benchmarks and A Chain of Summarization Approach}. In \bibinfo{booktitle}{\emph{Advances in Neural Information Processing Systems}}, \bibfield{editor}{\bibinfo{person}{A.~Globerson}, \bibinfo{person}{L.~Mackey}, \bibinfo{person}{D.~Belgrave}, \bibinfo{person}{A.~Fan}, \bibinfo{person}{U.~Paquet}, \bibinfo{person}{J.~Tomczak}, {and} \bibinfo{person}{C.~Zhang}} (Eds.), Vol.~\bibinfo{volume}{37}. \bibinfo{publisher}{Curran Associates, Inc.}, \bibinfo{pages}{133386--133442}.
\newblock
\urldef\tempurl%
\url{https://proceedings.neurips.cc/paper_files/paper/2024/file/f0ebc318e2df08360b2df559e81602e5-Paper-Conference.pdf}
\showURL{%
\tempurl}


\bibitem[Melhart et~al\mbox{.}(2025)]%
        {melhart2025largelanguagemodelscapture}
\bibfield{author}{\bibinfo{person}{David Melhart}, \bibinfo{person}{Matthew Barthet}, {and} \bibinfo{person}{Georgios~N. Yannakakis}.} \bibinfo{year}{2025}\natexlab{}.
\newblock \bibinfo{title}{Can Large Language Models Capture Video Game Engagement?}
\newblock
\newblock
\showeprint[arxiv]{2502.04379}~[cs.CV]
\urldef\tempurl%
\url{https://arxiv.org/abs/2502.04379}
\showURL{%
\tempurl}


\bibitem[microsoft(2025)]%
        {Azure}
\bibfield{author}{\bibinfo{person}{microsoft}.} \bibinfo{year}{2025}\natexlab{}.
\newblock \bibinfo{booktitle}{\emph{Azure}}.
\newblock
\urldef\tempurl%
\url{https://azure.microsoft.com/en-ca}
\showURL{%
\tempurl}


\bibitem[Mirza et~al\mbox{.}(2025)]%
        {meta_prompting_zero_shot}
\bibfield{author}{\bibinfo{person}{M.~Jehanzeb Mirza}, \bibinfo{person}{Leonid Karlinsky}, \bibinfo{person}{Wei Lin}, \bibinfo{person}{Sivan Doveh}, \bibinfo{person}{Jakub Micorek}, \bibinfo{person}{Mateusz Kozinski}, \bibinfo{person}{Hilde Kuehne}, {and} \bibinfo{person}{Horst Possegger}.} \bibinfo{year}{2025}\natexlab{}.
\newblock \showarticletitle{Meta-prompting for Automating Zero-Shot Visual Recognition with LLMs}. In \bibinfo{booktitle}{\emph{Computer Vision -- ECCV 2024}}, \bibfield{editor}{\bibinfo{person}{Ale{\v{s}} Leonardis}, \bibinfo{person}{Elisa Ricci}, \bibinfo{person}{Stefan Roth}, \bibinfo{person}{Olga Russakovsky}, \bibinfo{person}{Torsten Sattler}, {and} \bibinfo{person}{G{\"u}l Varol}} (Eds.). \bibinfo{publisher}{Springer Nature Switzerland}, \bibinfo{address}{Cham}, \bibinfo{pages}{370--387}.
\newblock
\showISBNx{978-3-031-72627-9}


\bibitem[Mistral(2025)]%
        {mistral}
\bibfield{author}{\bibinfo{person}{Mistral}.} \bibinfo{year}{2025}\natexlab{}.
\newblock \bibinfo{booktitle}{\emph{Frontier AI. In Your Hands}}.
\newblock
\urldef\tempurl%
\url{https://mistral.ai/}
\showURL{%
\tempurl}


\bibitem[Moran et~al\mbox{.}(2015)]%
        {Moran_2015}
\bibfield{author}{\bibinfo{person}{Kevin Moran}, \bibinfo{person}{Mario Linares-Vásquez}, \bibinfo{person}{Carlos Bernal-Cárdenas}, {and} \bibinfo{person}{Denys Poshyvanyk}.} \bibinfo{year}{2015}\natexlab{}.
\newblock \showarticletitle{Auto-completing bug reports for Android applications}. In \bibinfo{booktitle}{\emph{Proceedings of the 2015 10th Joint Meeting on Foundations of Software Engineering}}. \bibinfo{publisher}{ACM}, \bibinfo{pages}{673–686}.
\newblock
\urldef\tempurl%
\url{https://doi.org/10.1145/2786805.2786857}
\showDOI{\tempurl}


\bibitem[Nagar et~al\mbox{.}(2024)]%
        {zero_shot_Vlm}
\bibfield{author}{\bibinfo{person}{Aishik Nagar}, \bibinfo{person}{Shantanu Jaiswal}, {and} \bibinfo{person}{Cheston Tan}.} \bibinfo{year}{2024}\natexlab{}.
\newblock \showarticletitle{Zero-Shot Visual Reasoning by Vision-Language Models: Benchmarking and Analysis}. In \bibinfo{booktitle}{\emph{2024 International Joint Conference on Neural Networks (IJCNN)}}. \bibinfo{pages}{1--8}.
\newblock
\urldef\tempurl%
\url{https://doi.org/10.1109/IJCNN60899.2024.10650020}
\showDOI{\tempurl}


\bibitem[OpenAI(2024a)]%
        {GPT4O-mini}
\bibfield{author}{\bibinfo{person}{OpenAI}.} \bibinfo{year}{2024}\natexlab{a}.
\newblock \bibinfo{booktitle}{\emph{GPT-4o mini: advancing cost-efficient intelligence}}.
\newblock
\urldef\tempurl%
\url{https://openai.com/index/gpt-4o-mini-advancing-cost-efficient-intelligence/}
\showURL{%
\tempurl}


\bibitem[OpenAI(2024b)]%
        {GPT4O}
\bibfield{author}{\bibinfo{person}{OpenAI}.} \bibinfo{year}{2024}\natexlab{b}.
\newblock \bibinfo{booktitle}{\emph{Hello GPT-4o}}.
\newblock
\urldef\tempurl%
\url{https://openai.com/index/hello-gpt-4o/}
\showURL{%
\tempurl}


\bibitem[OpenAI(2024c)]%
        {GPT_API}
\bibfield{author}{\bibinfo{person}{OpenAI}.} \bibinfo{year}{2024}\natexlab{c}.
\newblock \bibinfo{booktitle}{\emph{OpenAI Platform}}.
\newblock
\urldef\tempurl%
\url{https://platform.openai.com/docs/api-reference/chat/create}
\showURL{%
\tempurl}


\bibitem[Paduraru et~al\mbox{.}(2021)]%
        {CV_auto_testing}
\bibfield{author}{\bibinfo{person}{Ciprian Paduraru}, \bibinfo{person}{Miruna Paduraru}, {and} \bibinfo{person}{Alin Stefanescu}.} \bibinfo{year}{2021}\natexlab{}.
\newblock \showarticletitle{Automated game testing using computer vision methods}. In \bibinfo{booktitle}{\emph{2021 36th IEEE/ACM International Conference on Automated Software Engineering Workshops (ASEW)}}. \bibinfo{pages}{65--72}.
\newblock
\urldef\tempurl%
\url{https://doi.org/10.1109/ASEW52652.2021.00024}
\showDOI{\tempurl}


\bibitem[Peeperkorn et~al\mbox{.}(2024)]%
        {peeperkorn2024temperaturecreativityparameterlarge}
\bibfield{author}{\bibinfo{person}{Max Peeperkorn}, \bibinfo{person}{Tom Kouwenhoven}, \bibinfo{person}{Dan Brown}, {and} \bibinfo{person}{Anna Jordanous}.} \bibinfo{year}{2024}\natexlab{}.
\newblock \bibinfo{title}{Is Temperature the Creativity Parameter of Large Language Models?}
\newblock
\newblock
\showeprint[arxiv]{2405.00492}~[cs.CL]
\urldef\tempurl%
\url{https://arxiv.org/abs/2405.00492}
\showURL{%
\tempurl}


\bibitem[Reddit(2022)]%
        {reddit_video_example}
\bibfield{author}{\bibinfo{person}{Reddit}.} \bibinfo{year}{2022}\natexlab{}.
\newblock \bibinfo{booktitle}{\emph{Visual bug when Wraith finisher gets interupted.}}
\newblock
\urldef\tempurl%
\url{https://www.reddit.com/r/apexlegends/comments/13i8gmp/visual_bug_when_wraith_finisher_gets_interupted}
\showURL{%
\tempurl}


\bibitem[Richardson(2010)]%
        {H264_encoder}
\bibfield{author}{\bibinfo{person}{Iain~E. Richardson}.} \bibinfo{year}{2010}\natexlab{}.
\newblock \bibinfo{booktitle}{\emph{The H.264 Advanced Video Compression Standard, Second Edition}}.
\newblock \bibinfo{publisher}{Wiley}.
\newblock


\bibitem[Senchenko et~al\mbox{.}(2022)]%
        {Senchenko_2022}
\bibfield{author}{\bibinfo{person}{Alexander Senchenko}, \bibinfo{person}{Naomi Patterson}, \bibinfo{person}{Hamman Samuel}, {and} \bibinfo{person}{Dan Ispir}.} \bibinfo{year}{2022}\natexlab{}.
\newblock \showarticletitle{SUPERNOVA: Automating Test Selection and Defect Prevention in AAA Video Games Using Risk Based Testing and Machine Learning}. In \bibinfo{booktitle}{\emph{2022 IEEE Conference on Software Testing, Verification and Validation (ICST)}}. \bibinfo{publisher}{IEEE}, \bibinfo{pages}{345–354}.
\newblock
\urldef\tempurl%
\url{https://doi.org/10.1109/icst53961.2022.00043}
\showDOI{\tempurl}


\bibitem[Taesiri and Bezemer(2025)]%
        {Taesiri_VideoGameBunny}
\bibfield{author}{\bibinfo{person}{Mohammad~Reza Taesiri} {and} \bibinfo{person}{Cor-Paul Bezemer}.} \bibinfo{year}{2025}\natexlab{}.
\newblock \showarticletitle{VIDEOGAMEBUNNY: Towards vision assistants for video games}. In \bibinfo{booktitle}{\emph{Proceedings of the IEEE/CVF Winter Conference on Applications of Computer Vision}} (2025-03-01).
\newblock


\bibitem[Taesiri et~al\mbox{.}(2024)]%
        {taesiri2024glitchbenchlargemultimodalmodels}
\bibfield{author}{\bibinfo{person}{Mohammad~Reza Taesiri}, \bibinfo{person}{Tianjun Feng}, \bibinfo{person}{Cor-Paul Bezemer}, {and} \bibinfo{person}{Anh Nguyen}.} \bibinfo{year}{2024}\natexlab{}.
\newblock \showarticletitle{GlitchBench: Can Large Multimodal Models Detect Video Game Glitches?}. In \bibinfo{booktitle}{\emph{Proceedings of the IEEE/CVF Conference on Computer Vision and Pattern Recognition (CVPR)}}. \bibinfo{pages}{22444--22455}.
\newblock


\bibitem[Taesiri et~al\mbox{.}(2025)]%
        {taesiri2025videogameqabenchevaluatingvisionlanguagemodels}
\bibfield{author}{\bibinfo{person}{Mohammad~Reza Taesiri}, \bibinfo{person}{Abhijay Ghildyal}, \bibinfo{person}{Saman Zadtootaghaj}, \bibinfo{person}{Nabajeet Barman}, {and} \bibinfo{person}{Cor-Paul Bezemer}.} \bibinfo{year}{2025}\natexlab{}.
\newblock \bibinfo{title}{VideoGameQA-Bench: Evaluating Vision-Language Models for Video Game Quality Assurance}.
\newblock
\newblock
\showeprint[arxiv]{2505.15952}~[cs.CV]
\urldef\tempurl%
\url{https://arxiv.org/abs/2505.15952}
\showURL{%
\tempurl}


\bibitem[Taesiri et~al\mbox{.}(2022)]%
        {taesiri2022clipmeetsgamephysicsbug}
\bibfield{author}{\bibinfo{person}{Mohammad~Reza Taesiri}, \bibinfo{person}{Finlay Macklon}, {and} \bibinfo{person}{Cor-Paul Bezemer}.} \bibinfo{year}{2022}\natexlab{}.
\newblock \showarticletitle{CLIP meets GamePhysics: towards bug identification in gameplay videos using zero-shot transfer learning}. In \bibinfo{booktitle}{\emph{Proceedings of the 19th International Conference on Mining Software Repositories}} (Pittsburgh, Pennsylvania) \emph{(\bibinfo{series}{MSR '22})}. \bibinfo{publisher}{Association for Computing Machinery}, \bibinfo{address}{New York, NY, USA}, \bibinfo{pages}{270–281}.
\newblock
\showISBNx{9781450393034}
\urldef\tempurl%
\url{https://doi.org/10.1145/3524842.3528438}
\showDOI{\tempurl}


\bibitem[Tan et~al\mbox{.}(2025)]%
        {tan2025imagerenhancingbugreport}
\bibfield{author}{\bibinfo{person}{Xuchen Tan}, \bibinfo{person}{Deenu Yadav}, \bibinfo{person}{Faiz Ahmed}, {and} \bibinfo{person}{Maleknaz Nayebi}.} \bibinfo{year}{2025}\natexlab{}.
\newblock \bibinfo{title}{ImageR: Enhancing Bug Report Clarity by Screenshots}.
\newblock
\newblock
\showeprint[arxiv]{2505.01925}~[cs.SE]
\urldef\tempurl%
\url{https://arxiv.org/abs/2505.01925}
\showURL{%
\tempurl}


\bibitem[Truelove et~al\mbox{.}(2023)]%
        {truelove2023findingneedlehaystackdetecting}
\bibfield{author}{\bibinfo{person}{Andrew Truelove}, \bibinfo{person}{Shiyue Rong}, \bibinfo{person}{Eduardo~Santana de Almeida}, {and} \bibinfo{person}{Iftekhar Ahmed}.} \bibinfo{year}{2023}\natexlab{}.
\newblock \bibinfo{title}{Finding the Needle in a Haystack: Detecting Bug Occurrences in Gameplay Videos}.
\newblock
\newblock
\showeprint[arxiv]{2311.10926}~[cs.SE]
\urldef\tempurl%
\url{https://arxiv.org/abs/2311.10926}
\showURL{%
\tempurl}


\bibitem[Vidler and Walsh(2025)]%
        {vidler2025playinggameslargelanguage}
\bibfield{author}{\bibinfo{person}{Alicia Vidler} {and} \bibinfo{person}{Toby Walsh}.} \bibinfo{year}{2025}\natexlab{}.
\newblock \bibinfo{title}{Playing games with Large language models: Randomness and strategy}.
\newblock
\newblock
\showeprint[arxiv]{2503.02582}~[cs.AI]
\urldef\tempurl%
\url{https://arxiv.org/abs/2503.02582}
\showURL{%
\tempurl}


\bibitem[Wang et~al\mbox{.}(2025)]%
        {wang2025empiricalstudyleveragingimages}
\bibfield{author}{\bibinfo{person}{Dingbang Wang}, \bibinfo{person}{Zhaoxu Zhang}, \bibinfo{person}{Sidong Feng}, \bibinfo{person}{William G.~J. Halfond}, {and} \bibinfo{person}{Tingting Yu}.} \bibinfo{year}{2025}\natexlab{}.
\newblock \showarticletitle{An Empirical Study on Leveraging Images in Automated Bug Report Reproduction}. In \bibinfo{booktitle}{\emph{2025 IEEE/ACM 22nd International Conference on Mining Software Repositories (MSR)}}. \bibinfo{pages}{27--38}.
\newblock
\urldef\tempurl%
\url{https://doi.org/10.1109/MSR66628.2025.00019}
\showDOI{\tempurl}


\bibitem[Wilcoxon(1992)]%
        {Wilcoxon1992}
\bibfield{author}{\bibinfo{person}{Frank Wilcoxon}.} \bibinfo{year}{1992}\natexlab{}.
\newblock \bibinfo{booktitle}{\emph{Individual Comparisons by Ranking Methods}}.
\newblock \bibinfo{publisher}{Springer New York}, \bibinfo{address}{New York, NY}, \bibinfo{pages}{196--202}.
\newblock
\showISBNx{978-1-4612-4380-9}
\urldef\tempurl%
\url{https://doi.org/10.1007/978-1-4612-4380-9_16}
\showDOI{\tempurl}


\bibitem[Yan et~al\mbox{.}(2024)]%
        {yan2024semanticguiscenelearning}
\bibfield{author}{\bibinfo{person}{Yanfu Yan}, \bibinfo{person}{Nathan Cooper}, \bibinfo{person}{Oscar Chaparro}, \bibinfo{person}{Kevin Moran}, {and} \bibinfo{person}{Denys Poshyvanyk}.} \bibinfo{year}{2024}\natexlab{}.
\newblock \showarticletitle{Semantic GUI Scene Learning and Video Alignment for Detecting Duplicate Video-based Bug Reports}. In \bibinfo{booktitle}{\emph{Proceedings of the IEEE/ACM 46th International Conference on Software Engineering}} (Lisbon, Portugal) \emph{(\bibinfo{series}{ICSE '24})}. \bibinfo{publisher}{Association for Computing Machinery}, \bibinfo{address}{New York, NY, USA}, Article \bibinfo{articleno}{232}, \bibinfo{numpages}{13}~pages.
\newblock
\showISBNx{9798400702174}
\urldef\tempurl%
\url{https://doi.org/10.1145/3597503.3639163}
\showDOI{\tempurl}


\bibitem[Zhao et~al\mbox{.}(2021)]%
        {zhao2021lightweightapproachhumanlikeplaytesting}
\bibfield{author}{\bibinfo{person}{Yan Zhao}, \bibinfo{person}{Weihao Zhang}, \bibinfo{person}{Enyi Tang}, \bibinfo{person}{Haipeng Cai}, \bibinfo{person}{Xi Guo}, {and} \bibinfo{person}{Na Meng}.} \bibinfo{year}{2021}\natexlab{}.
\newblock \bibinfo{title}{A Lightweight Approach of Human-Like Playtesting}.
\newblock
\newblock
\showeprint[arxiv]{2102.13026}~[cs.SE]
\urldef\tempurl%
\url{https://arxiv.org/abs/2102.13026}
\showURL{%
\tempurl}


\bibitem[Zheng et~al\mbox{.}(2023)]%
        {zheng2023judgingllmasajudgemtbenchchatbot}
\bibfield{author}{\bibinfo{person}{Lianmin Zheng}, \bibinfo{person}{Wei-Lin Chiang}, \bibinfo{person}{Ying Sheng}, \bibinfo{person}{Siyuan Zhuang}, \bibinfo{person}{Zhanghao Wu}, \bibinfo{person}{Yonghao Zhuang}, \bibinfo{person}{Zi Lin}, \bibinfo{person}{Zhuohan Li}, \bibinfo{person}{Dacheng Li}, \bibinfo{person}{Eric Xing}, \bibinfo{person}{Hao Zhang}, \bibinfo{person}{Joseph~E Gonzalez}, {and} \bibinfo{person}{Ion Stoica}.} \bibinfo{year}{2023}\natexlab{}.
\newblock \showarticletitle{Judging LLM-as-a-Judge with MT-Bench and Chatbot Arena}. In \bibinfo{booktitle}{\emph{Advances in Neural Information Processing Systems}}, \bibfield{editor}{\bibinfo{person}{A.~Oh}, \bibinfo{person}{T.~Naumann}, \bibinfo{person}{A.~Globerson}, \bibinfo{person}{K.~Saenko}, \bibinfo{person}{M.~Hardt}, {and} \bibinfo{person}{S.~Levine}} (Eds.), Vol.~\bibinfo{volume}{36}. \bibinfo{publisher}{Curran Associates, Inc.}, \bibinfo{pages}{46595--46623}.
\newblock
\urldef\tempurl%
\url{https://proceedings.neurips.cc/paper_files/paper/2023/file/91f18a1287b398d378ef22505bf41832-Paper-Datasets_and_Benchmarks.pdf}
\showURL{%
\tempurl}


\end{thebibliography}
\end{document}